\documentclass[fleqn,usenatbib]{mnras}

\usepackage{newtxtext,newtxmath}

\usepackage[T1]{fontenc}

\DeclareRobustCommand{\VAN}[3]{#2}
\let\VANthebibliography\thebibliography
\def\thebibliography{\DeclareRobustCommand{\VAN}[3]{##3}\VANthebibliography}

\usepackage{graphicx}	
\usepackage{amsmath}	
\usepackage{amssymb}	

\title[dEB SB2 TV Mon]{TV Mon - post mass transfer Algol type binary with $\delta$ Scuti pulsations in primary component. }

\author[M. Kovalev et al.]{
Mikhail Kovalev,$^{1,2,3}$\thanks{E-mail: mikhail.kovalev@ynao.ac.cn}
Zhenwei Li,$^{1,2}$
Jianping Xiong,$^{1,2}$
Azizbek Matekov,$^{1,4,5}$
\newauthor
Zhang {Bo},$^{5}$
Xuefei Chen$^{1,2,6}$
and Zhanwen Han$^{1,2,6}$\\
$^{1}$Yunnan Observatories, China Academy of Sciences, Kunming 650216, China\\
$^{2}$Key Laboratory for the Structure and Evolution of Celestial Objects, Chinese Academy of Sciences, Kunming 650011, China\\
$^{3}$International Centre of Supernovae, Yunnan Key Laboratory, Kunming 650216, China\\
$^{4}$ University of Chinese Academy of Sciences, Yuquan Road 19, Beijing 100049 Sijingshang Block, China\\
$^{5}$Ulugh Beg Astronomical Institute, Uzbekistan Academy of Sciences, 33 Astronomicheskaya str., Tashkent, 100052, Uzbekistan\\
$^{6}$Key Laboratory of Optical Astronomy, National Astronomical Observatories, Chinese Academy of Sciences, Beijing 100101, China\\
$^{7}$Center for Astronomical Mega-Science, Chinese Academy of Sciences, 20A Datun Road, Chaoyang District, Beijing 100012, China\\
}

\date{Accepted XXX. Received YYY; in original form ZZZ}

\pubyear{2024}

\def\kms{\,{\rm km}\,{\rm s}^{-1}}

\def\feh{\hbox{[Fe/H]}}

\newcommand{\teff}{T_{\rm eff}}
\newcommand{\rv}{{\rm RV}}
\def\Vmic{V_{\rm mic}}

\def\vsini{V \sin{i}}
\def\logg{\log{\rm (g)}}
\def\snr{\hbox{S/N}}
\newcommand{\ha}{\hbox{H$\alpha$}}

\begin{document}
\label{firstpage}
\pagerange{\pageref{firstpage}--\pageref{lastpage}}
\maketitle

\begin{abstract}
We present a study of the detached eclipsing binary TV~Mon using spectra from the LAMOST medium-resolution survey and ASAS-SN, CoRoT photometry.
We apply multiple-epochs spectral fitting to derive RV and spectral parameters. The analysis of eclipses in CoRoT data show the relative sizes of the stellar components and almost edge-on circular orbit. Combining the spectral and photometrical solutions we estimate masses and radii of the components:  $M_{A,B}=2.063\pm0.033({\rm stat.})\pm0.095({\rm syst.}),~0.218\pm0.004({\rm stat.})\pm0.018({\rm syst.})~M_\odot$, $R_{A,B}=2.394\pm0.014,~2.860\pm0.016~R_\odot$. SED analysis and Gaia parallax allow us to get estimation of temperatures ${\teff}_{A,B}=7624^{+194}_{-174},~5184^{+130}_{-123}$ K and distance $d=907\pm11$ pc. We identify three $\delta$ Scuti type pulsation frequencies in the primary component, while we also suspect TV~Mon having a spot activity in the secondary component. This system experienced intensive mass transfer and mass ratio reversal in the past, but currently shows no signs of mass transfer in the spectra. The low mass component will lose its outer envelope and shrink to the helium white dwarf, the mass and orbital period of which are in good agreement with evolutionary models predictions. 

\end{abstract}




\section{Introduction}

Detached eclipsing binaries (dEBs) are very important astronomical objects, which allow us to study stellar structure and evolution \citep{torres2010,zw2020}. Many close eclipsing binaries have significantly different masses of the components, which indicates there was mass-transfer in the past \citep{chen2020}. Quite often, high-frequency pulsations can be found in these systems using precise space-based photometry \citep{chen_ooDra,oscillations}.   
Recently \cite{xiong} derived masses and radii for 56 dEBs by using the LAMOST (Large Sky Area Multi-Object fiber Spectroscopic Telescope, also known as Guoshoujing telescope) medium-resolution survey (MRS) spectra and ground-based time-domain photometry. We select one of these stars, TV Mon, due to it's large mass ratio and analyse it using precise space-based photometry and multiple-epoch spectra fitting, developed in \cite{tyc,j0647}.
\par
 TV Mon was discovered on photographic plates by Sergei Iwanovitsch Beljawsky and classified as Algol-type eclipsing binary \citep{firstref}. The catalogue by \cite{svechnikov1990} reported it as a system with semi-detached configuration. 
With a location close to the Rosette nebula, many studies consider it is being a member of the associated open cluster NGC~2237 \citep{memb1,memb2,memb3}. However the most recent analysis rejects membership with probability of association $\sim0.018$ \citep{rosette_nebula}. 
\par
The paper is organised as follows: in Sections~\ref{sec:obs} and \ref{sec:methods} we describe the observations and methods. Section~\ref{results} presents our results. In Section~\ref{discus} we discuss the results in context of binary system evolution. In Section~\ref{concl} we summarise the paper and draw conclusions.

\section{Observations}
\label{sec:obs}
\subsection{Spectra}

LAMOST is a 4-meter quasi-meridian reflective Schmidt telescope with 4000 fibers installed on its $5\degr$ field of view focal plane. These configurations allow it to observe spectra for at most 4000 celestial objects simultaneously \citep{2012RAA....12.1197C, 2012RAA....12..723Z}.
 All available spectra were downloaded from \url{www.lamost.org/dr9/} under the designation J062822.72+051257.2.	We use the spectra taken at a resolving power of R$=\lambda/ \Delta \lambda \sim 7\,500$. Each spectrum is divided into two arms: blue from 4950\,\AA~to 5350\,\AA~and red from 6300\,\AA~to 6800\,\AA. We convert the heliocentric wavelength scale in the observed spectra from vacuum to air using {\sc PyAstronomy} \citep{pya} and apply zero point correction using values from \cite{zb_rv}, when it is possible. Observations are carried out from 2019-02-16 till 2020-03-05, covering 12 nights with time base of 373 days.
 The period is short ($P=4.179742$ d \cite{period_ref1}) therefore we analysed spectra taken during short 20 minutes exposures, unlike \cite{tyc,cat22}, where spectra stacked for the whole night were used.  In total we have 35 spectra, where the average signal-to-noise ratio ($\snr$) of a spectrum ranges from 9 to 60 ${\rm pix}^{-1}$ for the blue arm and from 10 to 82 ${\rm pix}^{-1}$ for the red arm of the spectrum, with the majority of the spectra having $\snr$ around 40 ${\rm pix}^{-1}$.

\subsection{Photometry}

We extracted publicly available $g$ and $V$ band light curves (LCs) from the ASAS-SN portal\footnote{\url{https://asas-sn.osu.edu/variables/243496}}. They contain 598 and 4396 datapoints respectively and covers a timebase of 3596 days.     
\par
\begin{figure*}
    \includegraphics[width=\textwidth]{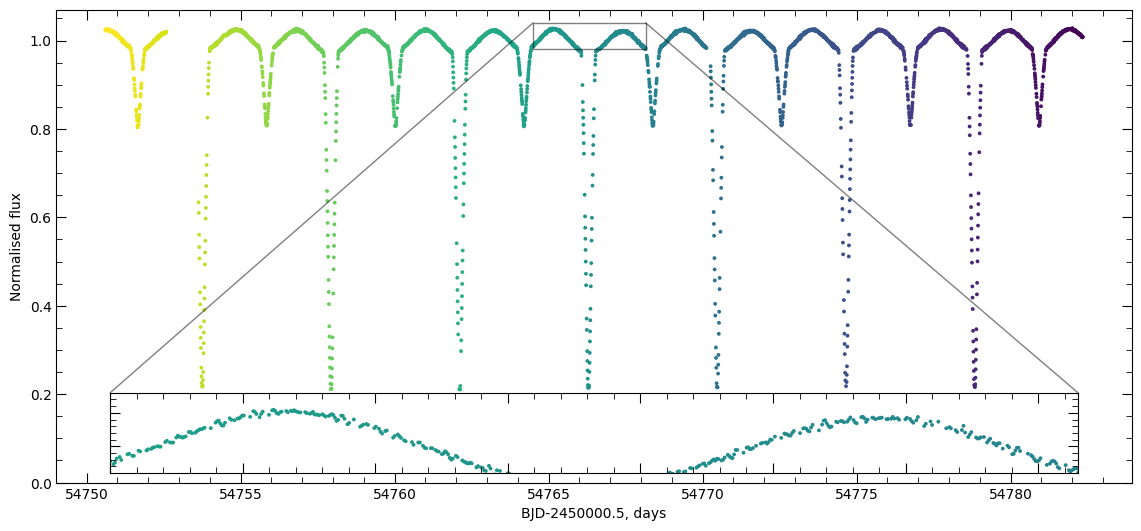}
    \includegraphics[width=\textwidth]{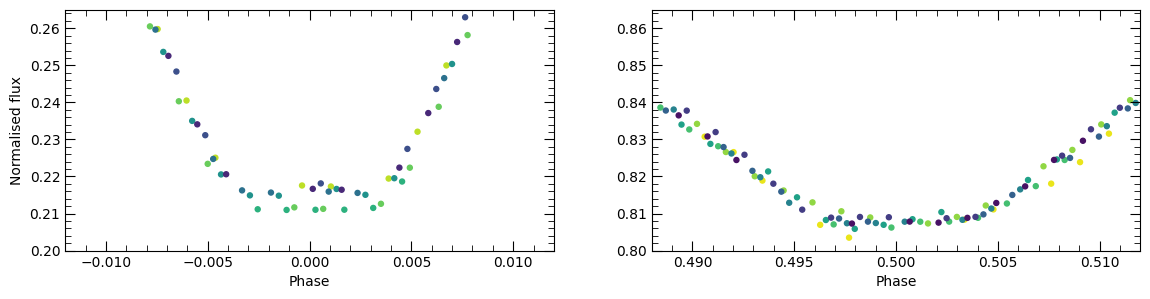}
    \caption{Normalised CoRoT light curve, with inline plot showing gaps and high-frequency pulsations (top panel) and phased eclipses, showing flat regions (bottom panel).}
    \label{fig:tess2}
\end{figure*}

The CoRoT (Convection, Rotation and planetary Transit) satellite mission\citep{corot} provides very high-quality LCs, due to lack of atmospheric noise and high sensitivity. We download it\footnote{available on VizieR~\url{http://cdsarc.u-strasbg.fr/saadavizier/download?oid=1153203439145087791} \citep{corot_data}} and use systematic-corrected ``WHITEFLUXSYS" data, with ``STATUS" field equals to zero (3443 datapoints in total, time coverage of $\sim32$ days). We normalise this dataset by dividing it by median values in the out-of-eclipse regions of the two time intervals: BJD<2454753.5 days (exposure time $512^{s}$) and BJD>2454753.5 days (exposure time $32^{s}$). The phase-folded LC shows two eclipses with short, flat minima ($\sim40$ min), while many small gaps and high-frequency pulsations are clearly visible, see Figure~\ref{fig:tess2}. These gaps are artefacts of removal of the ``jumps", caused by the cosmic radiation, which mostly occurred when the satellite crossed the South Atlantic Anomaly (SAA) \citep{corot1,corot_eb}.  \par
Timeseries from the {\sc WISE} satellite\citep{wise} are available for this system in $W1$ and $W2$ filters, see Appendix~\ref{sec:wise}. In infrared light the secondary eclipse is much deeper than in visual, which indicates a larger contribution of the secondary component to the total light of the system in this regime. However these LCs contain only 78 datapoints each and limb darkening coefficients are unavailable for these filters, so we don't use these data for the LC analysis. 
\par
Unfortunately the Transiting Exoplanet Survey Satellite \citep[TESS][]{tess} mission did not observe this system yet, because it is located close to the ecliptic plane, exactly in the gap between two sectors of observations. However the TESS input catalogue \citep{tic} contains it as TIC 234726486, so one can expect a new observations in the future. 

\section{Methods} 
\label{sec:methods}

\subsection{Spectral fitting}
\label{sec:maths} 

Our spectroscopic analysis includes two consecutive stages: 
\begin{enumerate}
    \item analysis of individual observations by binary and single-star spectral models, where we normalise the spectra and make a rough estimation of the spectral parameters, see brief description in Section~\ref{sec:ind}. 
    \item simultaneous fitting of multiple-epochs with a binary spectral model, using constraints from binary dynamics and values from the previous stage as an input, see Section~\ref{sec:multi}.
\end{enumerate}

\subsubsection{Individual spectra.}
\label{sec:ind}
The single-star spectral model is the same as in \cite{cat23} and described in Appendix~\ref{sec:payne}.
The normalised binary model spectrum is generated as a sum of the two Doppler-shifted, normalised single-star model spectra ${f}_{\lambda,i}$ scaled according to the difference in luminosity, which is a function of the $\teff$ and stellar size. We use following equation:    

\begin{align}
    {f}_{\lambda,{\rm binary}}=\frac{{f}_{\lambda,2} + k_\lambda {f}_{\lambda,1}}{1+k_\lambda},~
    k_\lambda= \frac{B_\lambda(\teff{_{,1}})}{B_\lambda(\teff{_{,2}})} k_R,
	\label{eq:bolzmann}
\end{align}
 where  $k_\lambda$ is the luminosity ratio per wavelength unit, $B_\lambda$ is the black-body radiation  (Plank function), $\teff$ is the effective temperature, $k_R$ - light ratio coefficient. Throughout the paper we always assume the primary star to be brighter.
\par
The binary model spectrum is later multiplied by the normalisation function, which is a linear combination of the first four Chebyshev polynomials \citep[similar to][]{Kovalev19}, defined separately for blue and red arms of the spectrum. The resulting spectrum is compared with the observed one using the \texttt{scipy.optimise.curve\_fit} function, which provides optimal spectral parameters, radial velocities (RV) of each component plus the light ratio and two sets of four coefficients of Chebyshev polynomials. We keep metallicity equal for both components. In total we have 18 free parameters for a binary fit.

\subsubsection{Multiple-epochs fitting.}
\label{sec:multi}

We explore the results from the fitting of the individual epochs and find that a result's quality clearly depends on separation of RVs. Clear double-lined spectra show that spectral lines are significantly broadened ($\vsini \sim 30~\kms$), thus the Rossiter-McLaughlin (RM) effect \citep{rossiter,mclaflin} can be observed during eclipses. 
\par
If two components in our binary system are gravitationaly bound, their radial velocities should agree with the following equation:
\begin{align}
\label{eqn:asgn}
    {\rm RV_{A}}=\gamma (1+q) - q {\rm RV_{B}},
\end{align}
where $q$ is the mass ratio  and $\gamma$ is the systemic velocity. Using this equation we can directly measure the systemic velocity and mass ratio. We should note that this equation is valid assuming the of absence of the RM effect, which can be justified by excluding spectral observations during the eclipses, and no difference in gravitational redshifts for two components $\Delta \gamma_i$. Given high mass ratio, we estimate this difference $\Delta\gamma_A-\Delta\gamma_B~\sim0.5~\kms$ using system parameters from \cite{xiong} and Equation 1 from \cite{grav_rs}, which is comparable to the usual precision of our $\rv$ measurements and the typical zero point offsets of LAMOST-MRS \citep{zb_rv}. Thus this effect can be neglected in our analysis.
\par
\par
To reduce the number of free parameters in the multiple-epochs fitting we can select only one component ${\rm RV}$ for fitting and compute values for another one using Equation~\ref{eqn:asgn}.  

We fit previously normalised individual epoch's spectra, using their binary spectral parameters values for initialisation. We repeat this five times using only manually selected best individual epoch's fit initialisation. We select the spectroscopic solution with minimal  $\chi^2$ as a final result. Surface gravity $\logg$ is poorly constrained by the spectra alone, thus we assign a value for $\logg$: initially we use to values from \cite{xiong}, and then to new estimation from the LC fit, see Section~\ref{sec:lc}. We repeat this iteratively until both spectra and LC were reasonably fitted.
\par
Finally we get new $\rv$ measurements for selected component, $q,~\gamma$ and spectral parameters for both components. We can repeat these calculations taking another component's RVs as a fitting parameters, but derived spectral parameters will be swapped and mass/light ratio will be inverted. Agreement between two solutions can validate assumptions of Equation~\ref{eqn:asgn}. Then we can use these RV measurements of both components in further analysis. 

\subsection{Light curve fitting}
\label{sec:lc}
Visual inspection of CoRoT LC indicates that there is a short total eclipse during the main minimum and a transit during the secondary minimum, separated by half of the period. Thus it is reasonable to assume an edge-on circular orbit with ``central" eclipse (inclination $i=90\degr$) and compute approximations for relative radii $R_i/a$ using pure geometry of spherical stars, see Formulae 12,13 from the great book by \cite{cesevich}:
\begin{align}
    R_A/a+R_B/a=\sin{\Theta_1},\\
    R_B/a-R_A/a=\sin{\Theta_2},\\
    2\Theta_1=D,~2\Theta_2=D_{\rm tot},
\end{align}
where $D$ and $D_{\rm tot}$ are durations of eclipse and it's total (flat) part, both divided by period, respectively, and $a$ is a radius of a circular orbit. This gives us $R_{A,B}/a\sim0.163,0.190$. These values can be used to initialize detailed LC modelling.  
\par 
We analyse the available LC datasets in two steps using different codes. Originally CoRoT LC and RV data were fitted by {\sc JKTEBOP} and Transit Light Curve Modeler code ({\sc TLCM}). These two codes assume a spherical shape of the stars and works extremely fast, although i is unable to fit several LC simultaneously. We use their solutions to check the consistency of our data and to derive period, mass ratio and orbital parameters of the system. Details about {\sc TLCM} modelling can be found in Appendix~\ref{sec:tlcm}. Subsequently the {\sc ellc} code was used to simultaneously fit ASAS-SN and CoRoT LCs with fixed period, mass ratio and orbit from the previous step.

\subsubsection{JKTEBOP}
We used the {\sc JKTEBOP} code (version 40)\footnote{{\sc jktebop} is written in {\sc fortran77} and the source code is available at \url{http://www.astro.keele.ac.uk/jkt/codes/jktebop.html}} by \citet{jkt} to simultaneously fit the LC and RV timeseries. We used quadratic limb darkening coefficients provided by the {\sc JKTLD} code\footnote{it is provided together with {\sc JKTEBOP}} and linearly interpolated them for the spectral parameters.
The eccentricity was set to zero and the systemic velocity was fitted for both binary components. In total we fit for 16 parameters: $J$ the central surface brightness ratio, $(R_A+R_B)/a$ the ratio of the sum of stellar radii to the semimajor axis, $R_B/R_A$ the ratio of the radii, $i$ the inclination, $S_0$ nuisance parameter for the out-of-eclipse magnitude, ${\rm reflected~light}_{A,B}$, $P$ the period, $t_0$, semiamplitudes and systemic velocities $K_{A,B}$, $\gamma_{A,B}$, and $S_0$. We use integration ring size $1^{\circ}$.  
\par
At first, we run the {\sc JKTEBOP} code in the mode ``Task 4'' to discard outliers larger than three sigma (only 26 datapoints were removed from LC dataset, while all 23 RV  were kept for both components) and allow it adjusting observational errors for LC and RV data, through several iterations until the reduced $\chi^2$ reaches unity.  Then we run {\sc JKTEBOP }in the Monte Carlo mode (Task 8 ``MC") to estimate uncertainties using 10000 simulations.

\subsubsection{ELLC}
The {\sc ellc } code \citep{ellc}, uses triaxial ellipsoids for each component's shape, which is better approximation than the sphere, especially when ``roche" shape is chosen. In order to save computational time we use only LC data from CoRoT and ASAS-SN without RV and fix $a\sin{i},~q,~P$ to the values from the previous solution. Limb and gravity darkening coefficients were computed using interpolator built-in to {\sc ellc}. Both LCs were fitted simultaneously with fractional radii $R_{A,B}/a$, inclination $i$, $t_0$, third light $L3_i$, surface brightness ratios for all datasets $J_g,~J_V,~J_C$\footnote{ {\sc ellc} have different definition of this parameter, than other two codes.} and ``heat" parameters $h_{A,B}$ which take into account reflection effects for both components.  Synchronicity parameters $F_{1,2}$ were set to 1, after several trial runs with different values, which couldn't reproduce flat minima. After a series of sets and trials we found out that the code is unable to fit flat parts of minima when whole CoRoT LC was used, thus we restrict our analysis only to eclipses from the time interval BJD=2454755.5:2454765.5 d (281 datapoints), which have less noise and shows clear flat minima with no signs of long-period variability, see Figure~\ref{fig:tess2}. The out-of-eclipse regions were fitted using ASAS-SN LCs, which are less affected by stellar pulsations, due to lower photometrical precision. Additionally, we found that the $h_A$ parameter was always zero, thus we fix it to this value. This is possibly due to insufficient precision of ASAS-SN LC to get an estimation of the very weak reflection effect. To get uncertainties we sample the solution using {\sc emcee} \citep{emcee} with 50 walkers and 2000 iterations.

\section{Results}
\label{results}

\begin{figure*}
	\includegraphics[width=\textwidth]{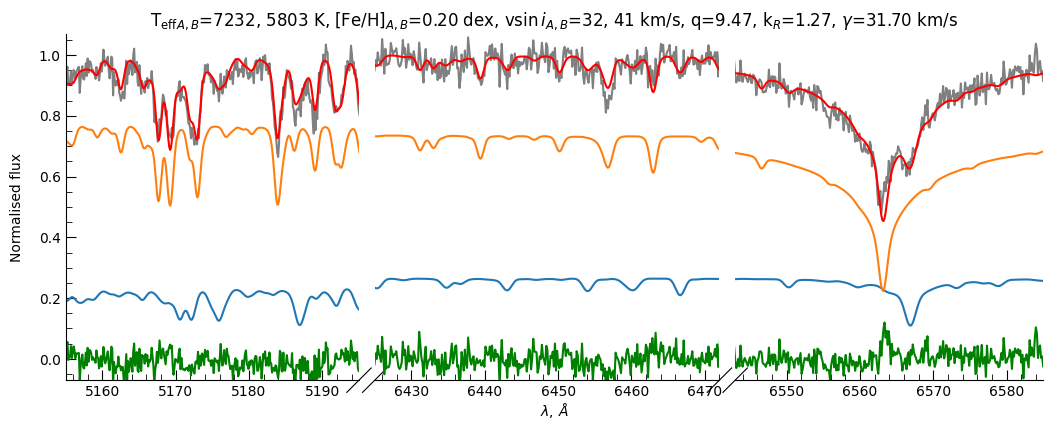}
	\includegraphics[width=\textwidth]{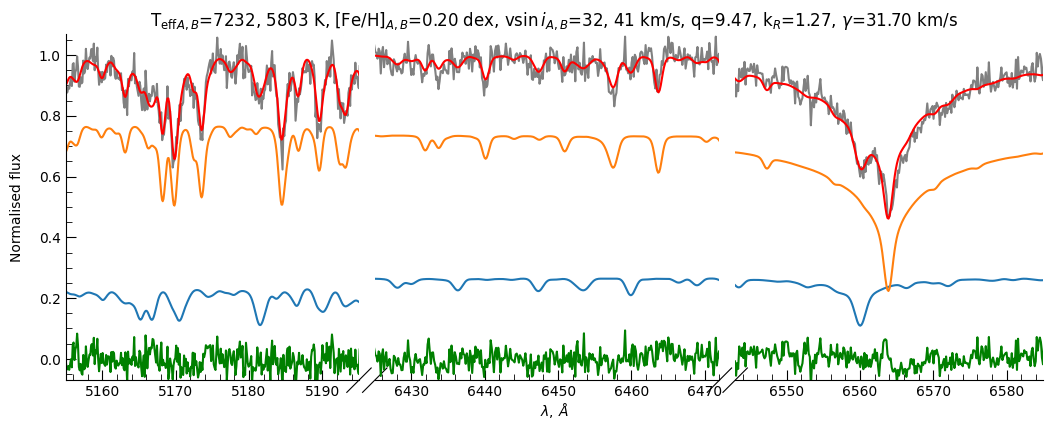}
    \caption{Example of the multiple-epoch fitting for spectra taken at $\phi=0.274$, MJD=58909.566 d (top) and $\phi=0.749$, MJD=58911.552 d (bottom). We zoom into the wavelength range around the magnesium triplet, $\ha$ and in a 70~\AA~interval in the red arm. The observed spectrum is shown as a gray line, the best fit is shown as red line. The primary component is shown as the orange line, the secondary as a blue line. The fit residuals are shown as a green line.}
    \label{fig:spfit}
\end{figure*}

In the Figure~\ref{fig:spfit} we show the best fit by multiple-epochs binary model for two epochs with large RV separation. We zoom into the wavelength range around the magnesium triplet, $\ha$ and in a 70~\AA~interval in the red arm, where many double lines are clearly visible. We can see that the primary star ($\teff=7232~K, \logg=3.96$ cgs, $\feh=0.2$ dex, $\vsini=32~\kms$) contributes around 75\% in the spectrum, while the secondary star ($\teff=5803~K, \logg=2.90$ cgs, $\feh=0.2$ dex, $\vsini=41~\kms$) contributes to the remaining 25\%. The derived mass ratio is $q\sim0.11$ and systemic radial velocity $\gamma=31.70~\kms$. Additionally we make another multiple-epoch fit with the secondary component's RV, finding the same spectral parameters with inverted mass ratio $1/q=9.47$. We present all spectroscopic results in the Table~\ref{tab:final}. The errors in the spectral parameters provided by \texttt{scipy.optimise.curve\_fit} are nominal and largely underestimated. Based on simulations with synthetic spectra in \cite{tyc,j0647}, typical errors of multiple-epoch spectral analysis are $\Delta{\teff}_{A,B}\sim 240,360~K,\,\Delta \logg_{A,B}=0.15,0.20$ cgs, $\Delta\feh=0.2$ dex, $\Delta \vsini_{A,B}=10,30~\kms$.  In principle, one can use spectroscopic $\vsini$ measurements to constrain the synchronicity parameters for LC modeling; however, our $\vsini$ can be an overestimate, since we assume rotation as only one source of line broadening. In reality one also needs to take into account the macroturbulence velocity \citep{gray}. 

\begin{table}
    \centering
    \caption{Spectral parameters from multiple-epoch fitting}
    \begin{tabular}{lcc}
\hline
Parameter & Star 1 & Star 2\\
\hline
$\gamma,\, \kms$ & \multicolumn{2}{c}{$31.70 \pm 0.10$}\\
$q$ & \multicolumn{2}{c}{$9.47\pm0.05$}\\
$k_R$  & \multicolumn{2}{c}{$1.27\pm0.02$}\\
$\teff$, K & 7232$\pm$3 & 5820$\pm$8 \\
fixed $\logg$, cgs & 3.96 & 2.90\\
$\feh$, dex & \multicolumn{2}{c}{$0.20\pm0.01$}\\
$\vsini,\,\kms$ & 32$\pm$1  & 41$\pm$1\\

\hline
    \end{tabular}
    \label{tab:final}
\end{table}

\par
\begin{figure}
	\includegraphics[width=\columnwidth]{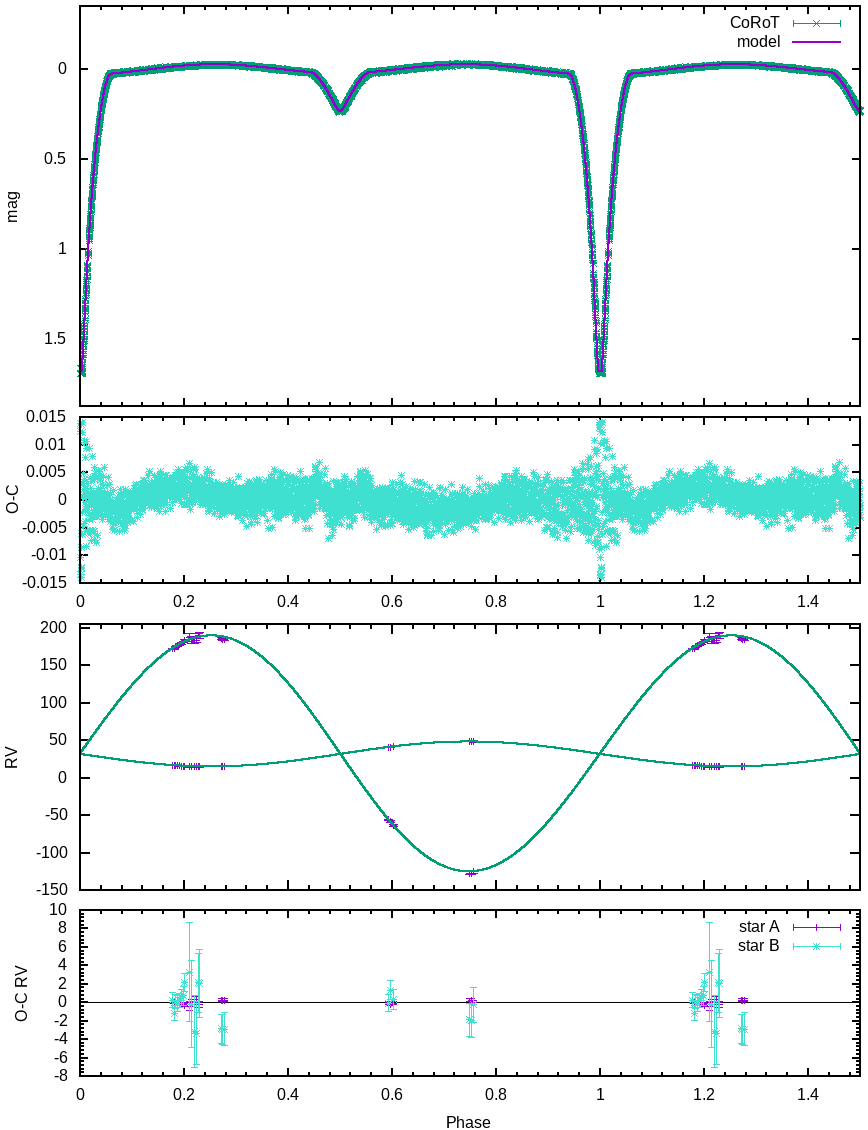}
    \caption{Phase-folded LC (top) and orbit (bottom) fits with {\sc JKTEBOP}.  The magnitudes are not calibrated.}
    \label{fig:corot_jkt}
\end{figure}

\begin{figure}
	\includegraphics[width=\columnwidth]{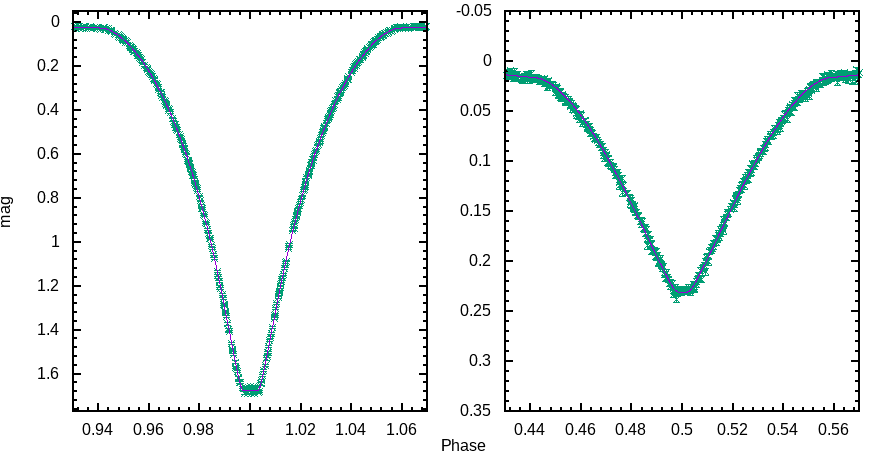}
    \caption{Same as Figure~\ref{fig:corot_jkt} for phases around eclipses. }
    \label{fig:zoom}
\end{figure}

   
In Figure~\ref{fig:corot_jkt} we show the best fits of the CoRoT LC (top) and RV (bottom) by {\sc JKTEBOP}, while the zoomed version which covers only eclipses can be found in Figure \ref{fig:zoom}. The fit residuals (O-C) are typically small: $\leq0.02$ mag in the total eclipse, $\leq 0.01$ mag for LCs outside of eclipse regions ( mostly due to high-frequency oscillations), and $\leq 4~\kms$ for RVs. Only seven measurements for $\rv_B$ are off by $\geq 1~\kms$ and correspond to relatively low $\snr$ spectra. The derived systemic velocities and mass ratio are similar to the values derived from a multiple-epoch spectral fit. Note $\gamma_A>\gamma_B$, as one would expect after taking into account the gravitational red shift, although this difference is well below uncertainties. We present results from {\sc JKTEBOP} in the Table~\ref{tab:lcorbit}.  
\par
The oblateness of the components ($0.0008, 0.0869$) is small, but for the secondary it is larger than the limit of $0.04$ \citep[][]{popper_etzel}, which can explain the relatively large residuals in total eclipse; therefore the {\sc JKTEBOP } solution should be supplemented by another code, for example {\sc ellc}. 

\par
There is another set of RV measurements from our spectra in \cite{zb_rv}, which we can use to check for possible systematic errors of our analysis. These RV were derived independently for the blue and red arms of the spectrum. We only used values derived from the blue arm of the spectra, because results for the red arm often are bad for the secondary component. We also excluded RV values from spectra taken during eclipses, to be consistent with multiple-epoch measurements. The results of {\sc JKTEBOP} for these RVs are presented in the right column of the Table~\ref{tab:lcorbit}. $K_{A,B}$ are slightly smaller than for the multiple-epochs solution, which propagates into the derived masses and sizes of components. The root mean square of the residuals is smaller for RVs from simultaneous fit of multiple-epochs, thus we prefer them for further analysis.

\begin{table*}
    \centering
        \caption{{\sc JKTEBOP} solutions using two RV datasets. Error estimates are from MC simulations.}
    \begin{tabular}{l|cc}
\hline
Parameter & CoRoT with multiple-epoch RV & CoRoT with \protect\cite{zb_rv} RV\\
\hline
fixed:\\
Grav. darkening$_A$ & \multicolumn{2}{c}{0.32}\\
Grav. darkening$_B$ & \multicolumn{2}{c}{0.32}\\
Limb darkening A &  \multicolumn{2}{c}{0.3527, 0.2934}\\
Limb darkening B &  \multicolumn{2}{c}{0.4548, 0.2264}\\
$e\cos{\omega}$  &  \multicolumn{2}{c}{0.0}\\
$e\sin{\omega}$  &  \multicolumn{2}{c}{0.0}\\

\hline
fitted:\\
$J$   & 0.2371$\pm$0.0010& 0.2371$\pm$0.0009 \\
$(R_A+R_B)/a$  &0.3644$\pm$0.0017 & 0.3644$\pm$0.0018 \\
$R_B/R_A$  & 1.1060$\pm$0.0022 & 1.1060$\pm$0.0022 \\
$i^{\circ}$  & 89.999$\pm$0.034 &89.999$\pm$0.034 \\
${\rm reflected~light}_A$, mag   & -0.003$\pm$0.004 &-0.003$\pm$0.004 \\
${\rm reflected~light}_B$, mag   & 0.002$\pm$0.004 &0.002$\pm$0.004 \\
$S_0$, mag & $-0.022\pm0.007$& $-0.022\pm0.007$ \\
$P$, d  & 4.179741$\pm$0.000011 &4.179729$\pm$0.000012 \\
$t_0$, BJD d& 2454457.495664$\pm$0.000210 & 2454457.496574$\pm$0.000348 \\
$K_A,\,\kms$  & 16.62$\pm$0.05 & 15.76$\pm$0.32 \\
$K_B,\,\kms$ & 157.38$\pm$0.46 & 155.32$\pm$0.55 \\
$\gamma_A,\,\kms$ & 31.72$\pm$0.09 & 31.78$\pm$0.29 \\
$\gamma_B,\,\kms$ & 31.56$\pm$0.81 & 31.45$\pm$0.79 \\
\hline
derived\\
$L_B/L_A$  & 0.318$\pm$0.017 &0.318$\pm$0.002 \\
$a,\,R_\odot$    & 14.375$\pm$0.040 & 14.134$\pm$0.055 \\
$q$   & 0.1056$\pm$0.0003 &0.1015$\pm$0.0020 \\
$M_A,\,M_\odot$    & 2.064$\pm$0.018 &1.969$\pm$0.022 \\
$M_B,\,M_\odot$    & 0.2179$\pm$0.0017 &0.200$\pm$0.005 \\
$R_A,\,R_\odot$    & 2.488$\pm$0.016 &2.446$\pm$0.016\\
$R_B,\,R_\odot$   & 2.751$\pm$0.016 & 2.705$\pm$0.016 \\
$\logg_A$, cgs   & 3.961$\pm$0.005 &3.955$\pm$0.005 \\
$\logg_B$, cgs   & 2.897$\pm$0.004 &2.874$\pm$0.010 \\
\hline
rms of the residuals:\\
LC, mmag& 5.230& 5.302\\
$\rv_{A},~\kms$ & 0.171 & 1.378\\
$\rv_{B},~\kms$ & 1.629& 1.287\\
\hline
    \end{tabular}
    \label{tab:lcorbit}
\end{table*}

\begin{figure}
    \centering
    \includegraphics[width=\columnwidth]{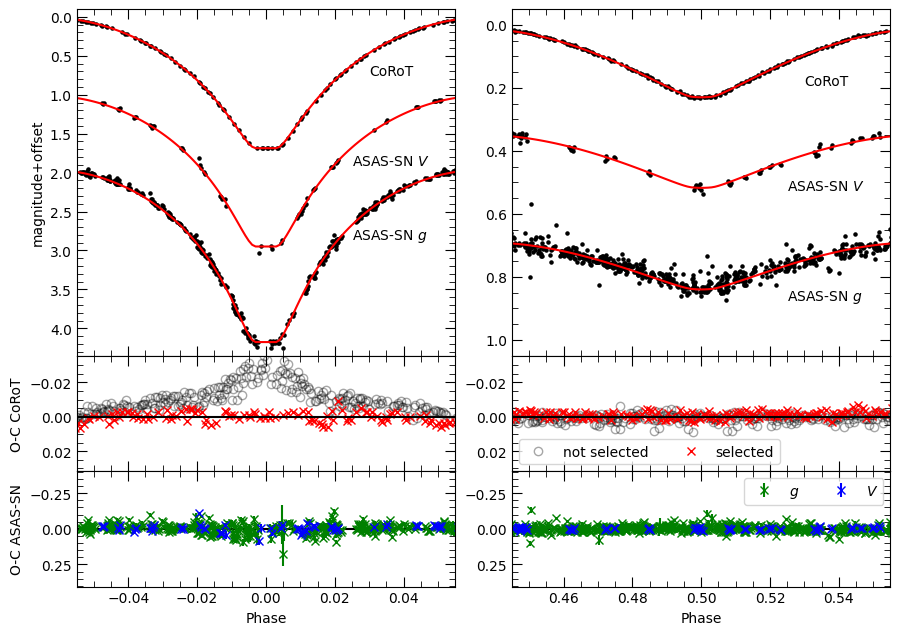}
    \includegraphics[width=\columnwidth]{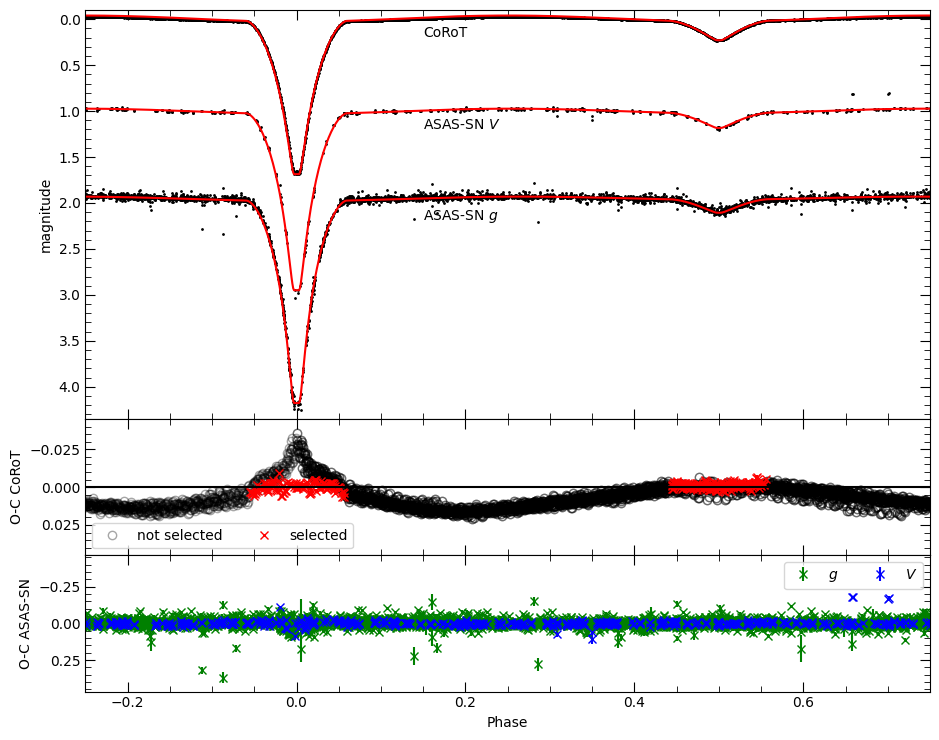}
    \caption{The LC solution from the {\sc ellc} code. We fit  whole ASAS-SN datasets, but only a part of CoRoT data from eclipses in the time interval BJD=2454755.5:2454765.5 was used in fitting (top panels). Full LCs are shown in bottom panel, while the fit residuals are shown below each panel. For the CoRoT data we also show the residuals computed for each datapoint, which were not used during fitting (gray/black open circles).}
    \label{fig:ecl_ellc}
\end{figure}

The results derived by {\sc ellc} are collected in Table~\ref{tab:ellc}. They are the median values and standard deviations for all fitted parameters. The related solution is depicted in Figure~\ref{fig:ecl_ellc}. This model provides an excellent fit for eclipses in CoRoT LC, typically resulting in residuals of $\leq0.01$ mag. The  ASAS-SN LCs are also well fitted, with typical residuals of $\leq0.05$ and $\leq0.04$ mag for $g$ and $V$ filters respectively.  If we compute fit residuals for the CoRoT datapoints outside the time interval BJD=2454755.5:2454756.5 d, they are significantly larger - up to 0.03 mag during the total eclipse, while there is almost no difference for the secondary minimum. This can indicate a slight long period variability of the secondary component (see Section~\ref{pulse}), which is not included in our LC model. The resulting values of radii differ from the previous solution primary being slightly smaller $R_A=2.39~R_\odot$ and the secondary is slightly larger $R_B=2.86~R_\odot$. The secondary radius is less than the critical value: $R_B/a=0.2020<R^{\rm lim}_{B}/a=0.2869$. The inclination is very similar to that in the {\sc JKTEBOP} solution, thus the masses are almost the same as in the previous solution, which provided us $a\sin{i}$ and $q$. The corner plot for the fitted parameters can be found in Figure~\ref{fig:corner}.

\begin{table}
    \centering
        \caption{ LC solutions by {\sc ellc} and {\sc W-D}. Error estimates are from median and standard deviations of {\sc emcee} sampling for {\sc ellc}, while for {\sc W-D} they are standard errors from the differential correction program.}
    \begin{tabular}{l|cc}
\hline
Parameter & {\sc ellc} value & {\sc W-D} value\\
\hline
fixed:\\
$P$, d  & \multicolumn{2}{c}{4.179741}\\
$a\sin{i},\,R_\odot$ & \multicolumn{2}{c} {14.375} \\
$q$ & \multicolumn{2}{c}{0.1056}\\
$e$ & \multicolumn{2}{c}{0.0}\\
$F_{1,2}$ & \multicolumn{2}{c}{1.0}\\
$h_A$ & 0.0 &\\
$A_1$ & & 1.0\\
${\teff}_A$, K & & 7232\\
\hline
fitted:\\
$R_A/a$  &0.16655$\pm$0.00009 & 0.1683\\
$R_B/a$  &0.19897$\pm$0.00008 & 0.1982\\
$J_g$   & 0.1117$\pm$0.0004& \\
$J_V$   & 0.1352$\pm$0.0012& \\
$J_C$   & 0.1883$\pm$0.0002& \\
$h_B$ & 0.351$\pm$0.004& \\
$t_0~g$, BMJD d& 54457.00198 $\pm$ 0.00001 & \\
$t_0~V$, BMJD d& 54456.99770 $\pm$ 0.00001 & \\
$t_0~C$, BMJD d& 54456.99566 $\pm$ 0.00001 & \\
$i^{\circ}$  & 89.36$\pm$0.01 & 89.52$\pm$0.08\\
${\teff}_B$, K & &$5054\pm3$\\
$L1_g$  & & 10.653$\pm$0.027\\
$L1_V$  & & 10.056$\pm$0.056\\
$L1_C$  & & 9.502$\pm$0.008\\
$L3_g$  & -0.0003$\pm$0.0006 & 0.001$\pm$0.002\\
$L3_V$  & 0.0115$\pm$0.0016 & 0.001$\pm$0.005\\
$L3_C$  & 0.0103$\pm$0.0003 & 0.025$\pm$0.001\\
$A_B$ & & 1.07$\pm$0.02 \\
$\Omega_A$ & & $6.059\pm0.009$\\
$\Omega_B$ & & $1.999\pm0.001$\\

\hline
derived\\%
$a,\,R_\odot$ & 14.375$\pm$0.076 & 14.375\\
$M_A,\,M_\odot$ & 2.063$\pm$0.033 & 2.063\\
$M_B,\,M_\odot$ & 0.218$\pm$0.004 & 0.218\\
$R_A,\,R_\odot$ & 2.394$\pm$0.014 & 2.419\\
$R_B,\,R_\odot$ & 2.860$\pm$0.016 & 2.849\\
$\logg_A$, cgs  & 3.994$\pm$0.002 & 3.99\\%
$\logg_B$, cgs  & 2.864$\pm$0.003 & 2.87\\
$r_1$ (pole) & & 0.1679$\pm$0.0003 \\
$r_1$ (point) & & 0.1685$\pm$0.0003 \\
$r_1$ (side) & & 0.1684$\pm$0.0003 \\
$r_1$ (back) & & 0.1685$\pm$0.0003 \\
$r_2$ (pole) & & 0.1856$\pm$0.0002 \\
$r_2$ (point) & & 0.2350$\pm$0.0005 \\
$r_2$ (side) & & 0.1921$\pm$0.0002 \\
$r_2$ (back) & & 0.2166$\pm$0.0003 \\
$\log{L_A},\,L_\odot$& & 1.157 \\
$\log{L_B},\,L_\odot$& & 0.676 \\
$L_B/L_A$ & & 0.295\\
\hline
    \end{tabular}
    \label{tab:ellc}
\end{table}

\subsection{Verification with {\sc W-D}}

\begin{figure}
    \centering
    \includegraphics[width=0.95\columnwidth]{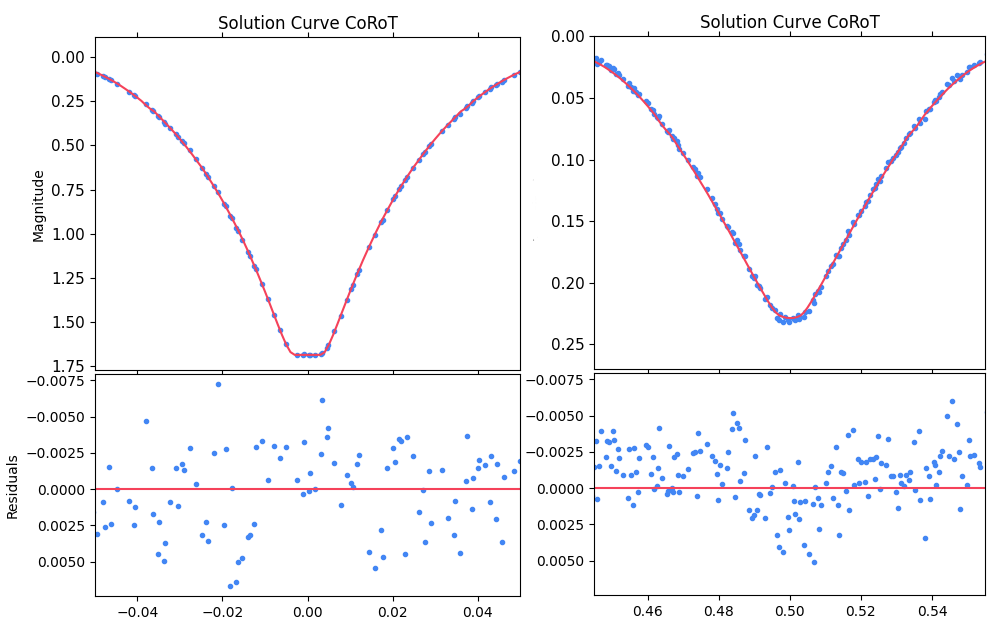}
    \includegraphics[width=0.95\columnwidth]{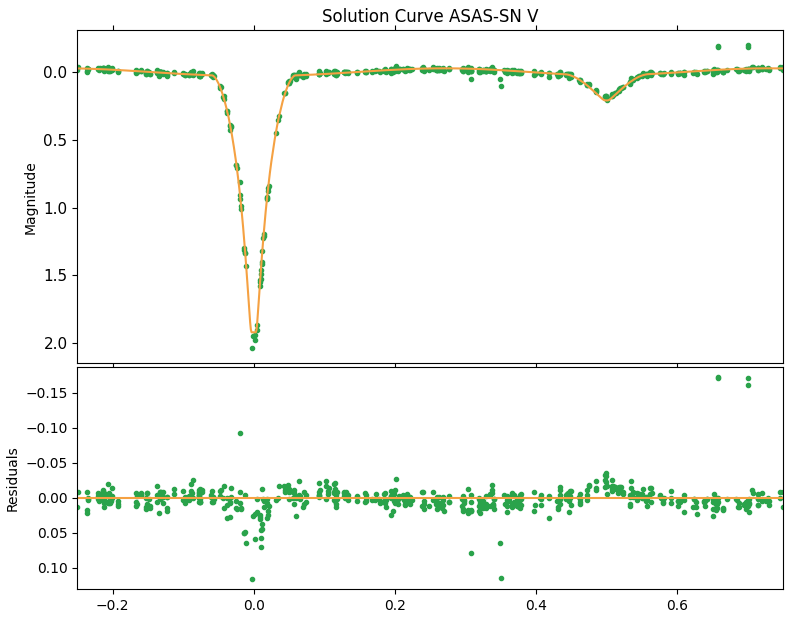}
    \includegraphics[width=0.95\columnwidth]{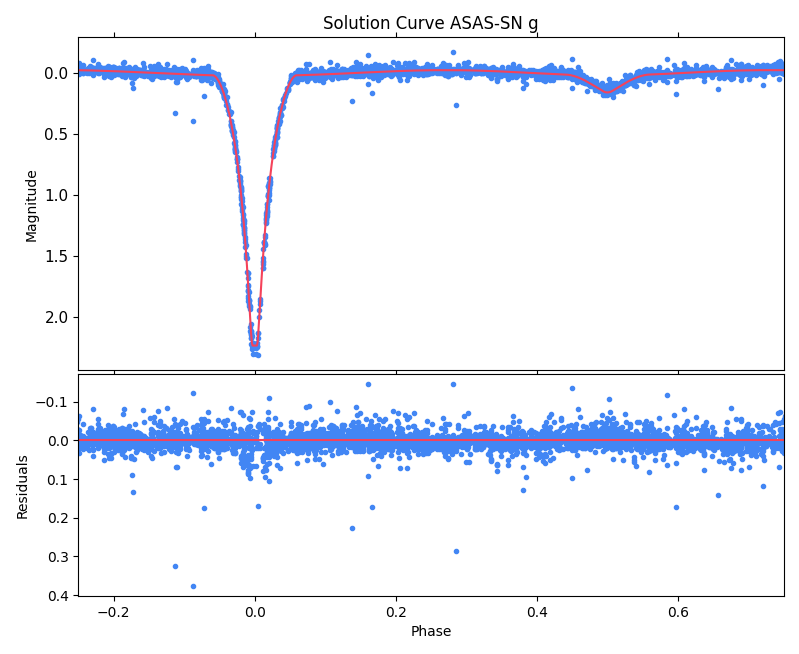}
    \caption{The LC solution by {\sc W-D} using the same data as the {\sc ellc} code. Top panels show fit of the data, bottom panels show fit residuals. }
    \label{fig:pywd}
\end{figure}

We use the Wilson–Devinney method {\sc W-D} \citep{wd71,wilson79} with {\sc PYWD2015} \citep{pywd} user interface to verify the {\sc ellc} solution. Same datasets were fitted by differential correction program with dimensionless potentials $\Omega$, inclination, ${\teff}_B$, the albedo of the secondary $A_B$ as free parameters. The bandpass luminosity $L1$ of the primary component and third light was also fitted for two LC, while the parameter $L2$ was computed by {\sc W-D}, based on the temperatures of the components. All other parameters were fixed, including the synchronicity parameters $F_{1,2}=1$. We show the solution derived after 21 iterations in Figure~\ref{fig:pywd} and Table~\ref{tab:ellc}. Residuals for the eclipses are $\leq5$ mmag for CoRoT dataset, ASAS-SN $g,~V$ mostly have residuals $\leq0.05$ mag, with larger residuals at the primary eclipse.  Overall the residuals are smaller than in {\sc ellc} possibly due to usage of the proper Roche geometry. The derived parameters are slightly different from {\sc ellc}, with the secondary component being a bit smaller and the primary is slightly larger. Also, one can get a good fit, only with non-zero third light contribution.   
\par
Since {\sc W-D} provides no errors for many parameters and the difference between {\sc ellc} and {\sc W-D} solutions is relatively small, we choose {\sc ellc} solution as a final result.  Additionally we run {\sc W-D} using all available data from the CoRoT dataset, together with both LCs from ASAS-SN finding similar results, see Appendix~\ref{sec:fullwd}.

\section{SED fitting}
\label{sec:sed}
\begin{figure*}
	\includegraphics[width=\textwidth]{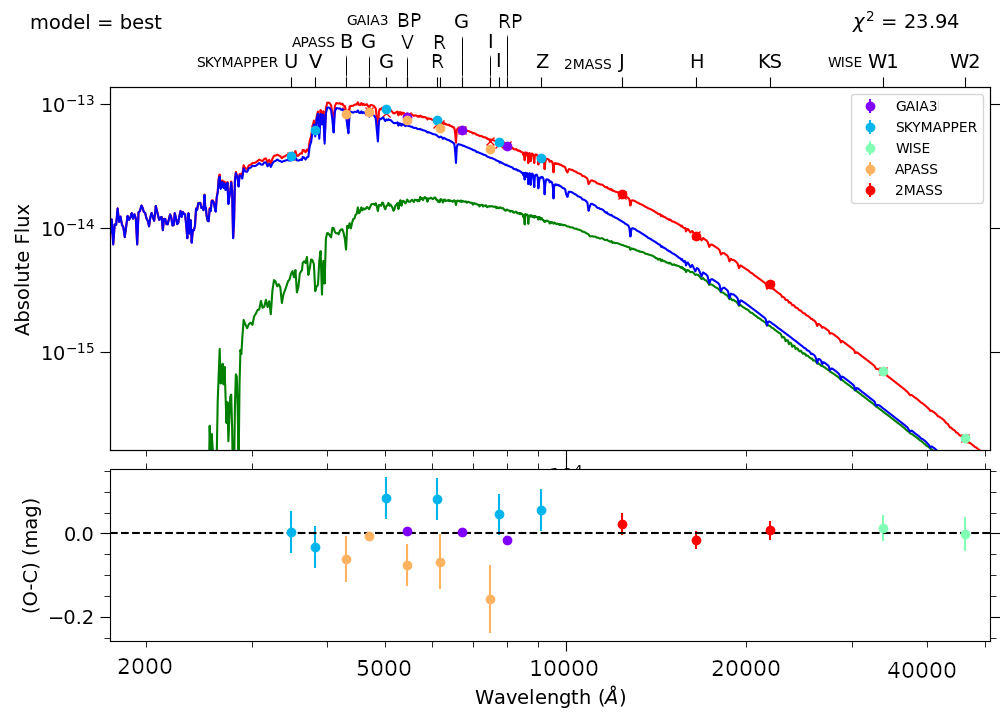}
    \caption{SED fitting with {\sc speedyfit}. The top panels shows the observations and spectral energy distribution of the system (red line), primary (blue line) and secondary (green line).  The fit residuals are shown on the bottom panel. }
    \label{fig:sed}
\end{figure*}

The spectral energy distribution (SED) offers an independent approach to estimating system parameters like $\teff$. We employ available photometry and aim to utilize out-of-eclipse data whenever feasible.
We utilize the {\sc speedyfit} package\footnote{https://github.com/vosjo/speedyfit} for SED fitting, using \cite{kurucz} models in conjunction with a Markov chain Monte Carlo method to identify the optimal fit and estimate the errors associated with the fitting parameters. We impose constraints on the solution by employing the parallax $\varpi=1.0654 \pm0.0201$ mas from Gaia DR3\citep{gaia3}, along with masses and radii derived from the LC solution. Errors in the constraints are included as priors in the Bayesian fitting process and are thus propagated to the final results. A more detailed description of the fitting process is given in \citet{Vos2017}. In Figure~\ref{fig:sed}, we present the resulting fit, while Table~\ref{tab:sed} lists the parameters. The estimated distance of $d=907\pm11$ pc is less than the single-star model value of $d=982^{1051}_{943}$ pc from Gaia DR3. The $\teff$ values deviate somewhat from the spectroscopic measurements, with the secondary component being cooler and the primary component slightly warmer in the SED fit. Most photometric measurements did not account for eclipse-related observations, apart from $W1$ and $W2$, which may explain the discrepancy, although alternative explanations could exist. 

\begin{table}
    \centering
        \caption{SED fitting results and observed photometry from Gaia DR3 \protect\cite{gaia3}, APASS \protect\cite{apass}, SKYMAPPER \protect\cite{skymapper}, 2MASS \protect\cite{2mass} and WISE \protect\cite{wise}. WISE values were computed as the mean and standard deviation of out-of-eclipse parts of the time-series.}
    \begin{tabular}{lc}
        \hline
        Observations (mag):\\
        GDR3 $G$ & $11.5146\pm0.0032$\\
        GDR3 $BP$ & $11.7826\pm0.0057$\\
        GDR3 $RP$ & $11.0990\pm0.0086$\\
        APASS $B$ & $12.19\pm0.06$\\
        APASS $V$ & $11.73\pm0.05$\\
        APASS $G$ & $11.878\pm0.006$\\
        APASS $R$ & $11.61\pm0.07$\\
        APASS $I$ & $11.58\pm0.08$\\
        SKYMAPPER $U$ & $13.41\pm0.05$\\
        SKYMAPPER $V$ & $12.68\pm0.05$\\
        SKYMAPPER $G$ & $11.71\pm0.05$\\
        SKYMAPPER $R$ & $11.50\pm0.05$\\
        SKYMAPPER $I$ & $11.42\pm0.05$\\
        SKYMAPPER $Z$ & $11.38\pm0.05$\\
        2MASS $J$ & $10.53\pm0.03$\\
        2MASS $H$ & $10.31\pm0.02$\\
        2MASS $K_s$ & $10.22\pm0.02$\\
        WISE $W1$ & $10.1678\pm0.0304$\\
        WISE $W2$ & $10.1983\pm0.0409$\\
        \hline
        Fitted parameters:\\
        $d$, pc& {$907_{-11}^{+11}$}\\
        $E(B-V)$, mag& {$0.170_{-0.027}^{+0.029}$}\\
        ${\teff}_{A,B}$, K& $7624_{-174}^{+194},\,5187_{-123}^{+130}$\\
        \hline
        
    \end{tabular}

    \label{tab:sed}
\end{table}

\section{Pulsations analysis}
\label{pulse}

\begin{figure*}
    \centering
    \includegraphics[width=\textwidth]{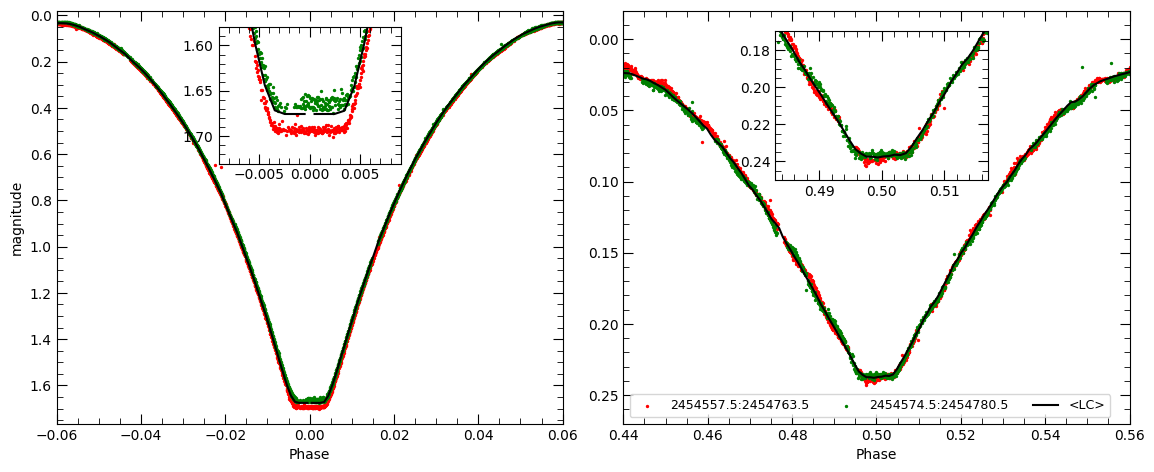}
    \caption{Mean LC and data for eclipses taken in the start and end parts of LC.}
    \label{fig:mean}
\end{figure*}

To explore pulsations in TV~Mon we use a different CoRoT photometry dataset ``WHITEFLUXFIL", available in the same datafile, which was not corrected for systematics, but has significantly more datapoints. These data have an exposure time of $32^{s}$ and contain 68103 datapoints with “STATUS" field equal to zero, which ensures that artifacts like ``jumps" due to crossing of SAA by satellite were excluded.  We normalise and flatten this LC by fitting a parabola to out-of-eclipse regions. Next, we use a similar approach to the \cite{chen_ooDra} analysis of OO~Dra. As our LC model unable to capture all possible variations of the LC for all time intervals, we computed an averaged curve for the phase-folded LC and then subtract it from the original LC. We use 301 phase bins, to ensure that each bin contains enough datapoints (typically 100-300). We show the mean LC and eclipses from the start and end of the original LC in Figure~\ref{fig:mean}. We see no pulsations in the primary eclipse, although it's flat part depth changes with time, which can indicate long period variability of the secondary component, due to fast changing spots or some instrumental effect in the CoRoT observations. The secondary eclipse shows clear pulsations, with the the flat part of the minima having similar depth. Thus, high-frequency pulsations come only from the primary component and LC data with phases from -0.06 till 0.06 were excluded from further analysis, leaving us with dataset of 60150 points.    

\begin{figure*}
    \centering
    \includegraphics[width=\textwidth]{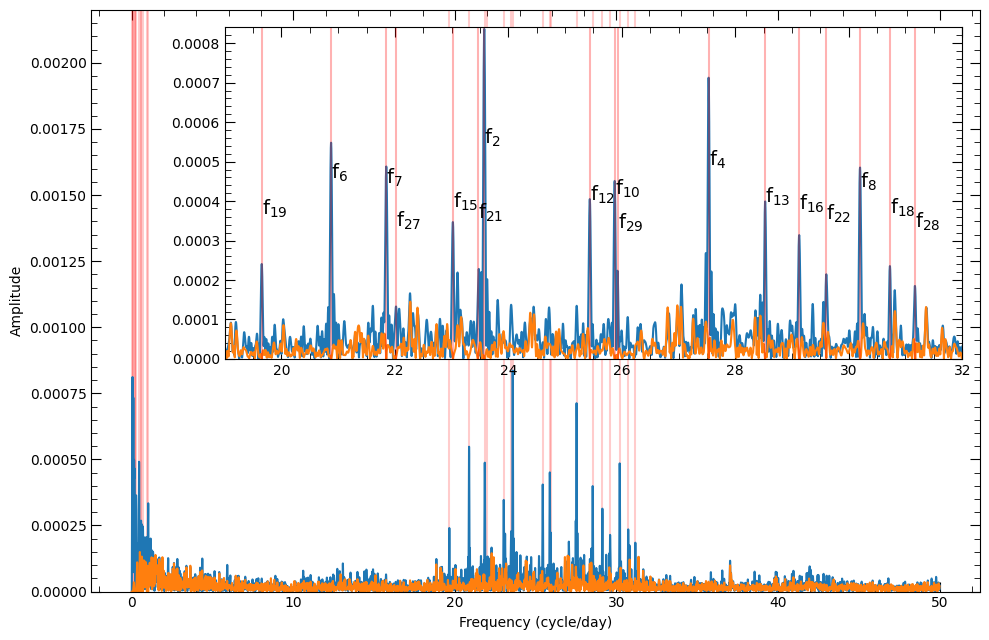}
    \includegraphics[width=\textwidth]{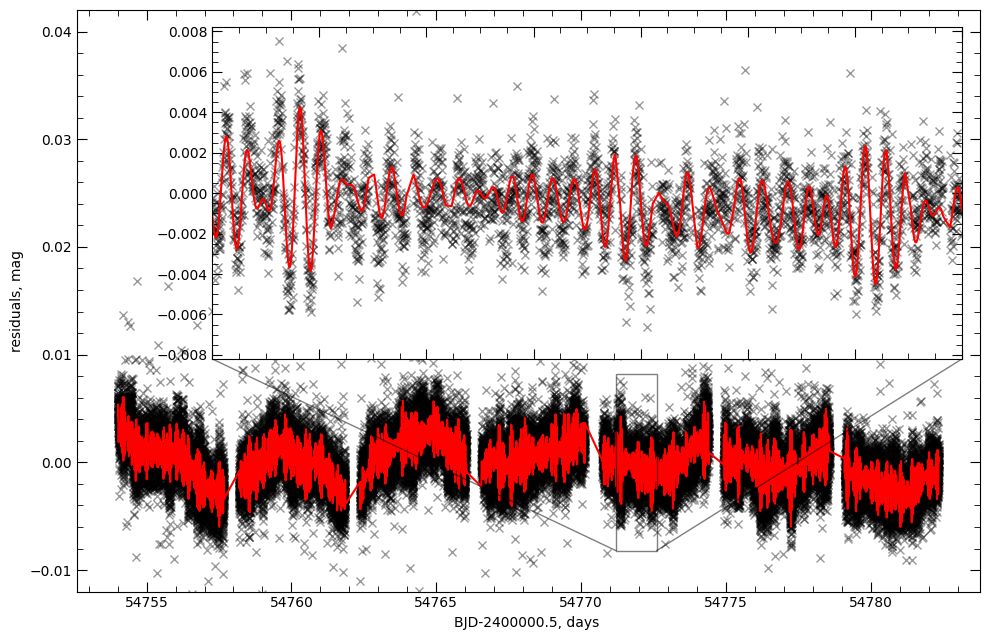}
    \caption{Fourier spectrum, with inline plot showing the region with possible $\delta$-Scuti pulsations (top) and best fit pulsation model (bottom). Vertical red lines indicate the position of all identified frequencies. The orange line shows the Fourier spectrum after the removal of 29 frequencies. }
    \label{fig:freq}
\end{figure*}

We explore pulsations using the {\sc Period04} software \citep{period04}. We follow the steps described in \cite{chen_ooDra} and \cite{oscillations} to iteratively find frequencies in the range from 0 to 50 cycles per day (no strong peaks were found with larger frequencies), until the residual Fourier spectrum has no peaks with $\snr>4$.  We identified 29 frequencies which are listed in Table~\ref{tab:freqs} and shown in the top panel of Figure~\ref{fig:freq}, while the best fit pulsations model is shown in the bottom panel. None of the frequencies match to $f_{\rm sat}=1/P_{\rm sat}=86400^{s}/6182^{s}=13.9715 ~{\rm day}^{-1}$  and it's multiples, where $P_{\rm sat}$ is the CoRoT satellite orbital period\citep{corot1}. Thus the CoRoT team successfully removed artifacts which might have arisen due to SAA crossing\footnote{As an exercise we computed Fourier spectrum for the raw "WHITEFLUX" minus mean LC. It shows two strong peaks with $f=f_{\rm sat},~4f_{\rm sat}$. }. With a frequency resolution of $0.055~{\rm day}^{-1}$\footnote{it is quite low due to short timebase of CoRoT observations} we also searched for possible orbital harmonics ($f_i=Nf_{\rm orb},\,f_{\rm orb}=1/P=0.239249\pm0.000001\,~{\rm day}^{-1}$) and combination frequencies. Unfortunately, many peaks (i.e. $f_1,f_3,f_5,f_9$) possibly originate from the imperfect removal of binarity-induced light variations, although some of them can be the result of instrumental effects or long period variability of the secondary component. For example, the period value $P_{17}\sim128$ d, corresponding to $f_{17}$, is similar to the long period $P_{\rm long}\sim30P_{\rm orb}$, detected in many Algol-type systems \citep{doubleperiod}.  Only $f_2=23.57,~f_4=27.53,~f_6=20.87~{\rm day}^{-1}$ can be considered as independent frequencies, possibly $\delta$ Scuti pressure modes. Although $f_5$ is close to  $115f_{\rm orb}$, it has a very strong peak in the Fourier spectrum and small error, which is enough to conclude that it is just an accidental agreement. Star A is likely to be a source of these pulsations, so we computed pulsation constants $Q=P_{\rm pul}\sqrt{\rho/\rho_\odot}$, where $P_{\rm pul}$ is the pulsation period, $\rho$ is the mean density of the star. We calculate $Q=0.014,\,0.016,\,0.018~{\rm day}$ for $f_4,f_2,f_6$ respectively. We leave detailed physical modelling of these pulsations (see an example in \cite{chen_ooDra}) to future studies. 

\begin{table*}
    \centering
\begin{tabular}[ht!]{lllll}
\hline
 &Frequency &Amplitude&Phase&comment\\ 
 & cycles per day & mmag & rad/(2$\pi$)&\\
\hline
independent frequencies:\\
$f_{2}$ &	23.5753$\pm$0.0002&	 0.8536$\pm$0.0089 &	 0.617$\pm$0.002 & $\delta {\rm Scuti}$\\
$f_{3}$ &	0.0605$\pm$0.0007&	 5.0481$\pm$0.1956 &	 0.139$\pm$0.003 & spot?\\
$f_{4}$ &	27.5301$\pm$0.0002&	 0.7048$\pm$0.0089 &	 0.069$\pm$0.002 & $\delta {\rm Scuti}$, $115f_{\rm orb}$\\
$f_{6}$ &	20.8750$\pm$0.0003&	 0.5569$\pm$0.0089 &	 0.042$\pm$0.003 & $\delta {\rm Scuti}$\\
$f_{17}$ &	0.0078$\pm$0.0016&	 1.4647$\pm$0.1523 &	 0.984$\pm$0.009 & long period?\\
combined frequencies and orbital harmonics:\\
$f_{1}$ &	0.2107$\pm$0.0002&	 1.3773$\pm$0.0112 &	 0.230$\pm$0.001 & $f_{\rm orb}$\\
$f_{5}$ &	0.0650$\pm$0.0006&	 4.7121$\pm$0.1950 &	 0.933$\pm$0.003 & $f_{3}$\\
$f_{7}$ &	21.8481$\pm$0.0003&	 0.5057$\pm$0.0089 &	 0.973$\pm$0.003 & $f_{6}+4 f_{\rm orb}$\\
$f_{8}$ &	30.1983$\pm$0.0003&	 0.4897$\pm$0.0089 &	 0.085$\pm$0.003 & $f_4+11 f_{\rm orb}$\\
$f_{9}$ &	0.1717$\pm$0.0005&	 0.5670$\pm$0.0102 &	 0.350$\pm$0.003 & $f_{\rm orb}$\\%
$f_{10}$ &	25.8730$\pm$0.0005&	 0.4164$\pm$0.0091 &	 0.300$\pm$0.003 & $f_4-7 f_{\rm orb}$\\
$f_{11}$ &	0.2687$\pm$0.0004&	 0.5643$\pm$0.0100 &	 0.991$\pm$0.003 & $f_3+f_{\rm orb}$\\
$f_{12}$ &	25.4367$\pm$0.0004&	 0.4357$\pm$0.0089 &	 0.725$\pm$0.003 & $f_7+12f_{\rm orb}$\\
$f_{13}$ &	28.5238$\pm$0.0004&	 0.3942$\pm$0.0089 &	 0.478$\pm$0.004 & $f_8-7 f_{\rm orb}$\\
$f_{14}$ &	0.4525$\pm$0.0005&	 0.4446$\pm$0.0099 &	 0.293$\pm$0.004 & $f_{11}+f_{\rm orb}$\\
$f_{15}$ &	23.0240$\pm$0.0005&	 0.3680$\pm$0.0089 &	 0.965$\pm$0.004 & $f_6+9 f_{\rm orb}$\\
$f_{16}$ &	29.1255$\pm$0.0005&	 0.3385$\pm$0.0089 &	 0.031$\pm$0.004 & $f_2+23 f_{\rm orb}$\\
$f_{18}$ &	30.7261$\pm$0.0007&	 0.2420$\pm$0.0089 &	 0.296$\pm$0.006 & $f_{12} +22 f_{\rm orb}$\\
$f_{19}$ &	19.6531$\pm$0.0007&	 0.2347$\pm$0.0088 &	 0.101$\pm$0.006 & $f_6-5 f_{\rm orb}$\\
$f_{20}$ &	0.5488$\pm$0.0020&	 0.2975$\pm$0.0354 &	 0.692$\pm$0.014 & $2 f_{11}$\\
$f_{21}$ &	23.4627$\pm$0.0008&	 0.2291$\pm$0.0089 &	 0.258$\pm$0.006 & $f_2-f_3$, side lobe?\\
$f_{22}$ &	29.6037$\pm$0.0007&	 0.2352$\pm$0.0089 &	 0.157$\pm$0.006 & $f_{14}+f_{16}$\\
$f_{23}$ &	0.5669$\pm$0.0018&	 0.3197$\pm$0.0359 &	 0.032$\pm$0.013 & $f_{20}$, side lobe?\\
$f_{24}$ &	0.6831$\pm$0.0008&	 0.2611$\pm$0.0096 &	 0.384$\pm$0.006 & $f_{14}+f_{\rm orb}$\\
$f_{25}$ &	1.0127$\pm$0.0009&	 0.2021$\pm$0.0090 &	 0.105$\pm$0.007 & $f_3+4 f_{\rm orb}$\\
$f_{26}$ &	0.9351$\pm$0.0009&	 0.1933$\pm$0.0097 &	 0.664$\pm$0.008 & $f_{14}+2 f_{\rm orb}$\\
$f_{27}$ &	22.0123$\pm$0.0011&	 0.1586$\pm$0.0089 &	 0.338$\pm$0.009 & $f_{21}-6 f_{\rm orb}$\\
$f_{28}$ &	31.1712$\pm$0.0011&	 0.1518$\pm$0.0089 &	 0.160$\pm$0.009 & $f_{12}+24 f_{\rm orb}$\\
$f_{29}$ &	25.9239$\pm$0.0012&	 0.1608$\pm$0.0092 &	 0.113$\pm$0.009 & $f_{10}$, side lobe \\
\hline 
\end{tabular}    
\caption{Pulsation frequencies, sorted based on their Fourier amplitudes. Errors are from {\sc Period04} based on \protect\cite{period04err}. }
    \label{tab:freqs}
\end{table*}

\section{Discussion}
\label{discus}
\subsection{System parameters}

This is an Algol-like eclipsing binary system, which experienced mass transfer and mass-ratio reversal in the past. The massive primary component is hot, while the much lighter secondary star is a cool red giant. Based on the absence of emission lines in $\ha$ region and secondary radius smaller than $R^{\rm lim}_B$ we can conclude that mass transfer is not active anymore.  
\par
We checked the ``O-C gateway" \citep{omc_gateway} to ensure stability of the period. It contains primary minima times from 1914 till 2019, based on visual, photographic and charge-coupled device (CCD) observations. When taking into account only the most accurate CCD based measurements, the period is constant, but a larger period ($P\sim4.17977~d$) is needed to fit earliest photographic measurements. If we trust these measurements, the period has slightly decreased and then become constant, although accuracy of the old photographic measurements can be very small and is hard to be evaluated now. Also TV Mon have total eclipse with duration $\sim40$ min, which can cause additional bias if not taken into account.
\par
Analysis of pulsations revealed three distinct frequencies of $\delta$ Scuti type associated with the primary component. Several low frequency pulsations were also identified, which can be interpreted as imperfect removal of binary flux variations or/and slow variability caused by spots. We checked VizieR for other CoRoT targets observed simultaneously with TV~Mon and found five in $2\arcmin$ cone. Two of them belong to spectral type K3II and show irregular variability, while two others have the same spectral type as TV~Mon (A5IV) and have stable LCs. Thus we think that slow variability is real and is not caused by some instrumental effect. The secondary component is a cool red giant, which can have magnetic activity, required to produce spots. As these spots seems to change very fast (see lower panel in Figure~\ref{fig:freq}), we don't model them in our LC solutions.         
\par
Previous work by \cite{xiong} used ASAS-SN photometry together with LAMOST-MRS RV from \cite{zb_rv}, while we utilise better photometry and updated RV, based on the same spectra. They estimate  $M_{A,B}=2.024\pm0.040,\,0.21\pm0.04,\,M_\odot$, $R_{A,B}=2.357\pm0.187,\,2.869\pm0.151,\,R_\odot$ .As we show in Section~\ref{results}, a systematic difference in RV will lead to slightly different stellar masses. This difference ($\sim5$ per cent) can be seen as a realistic estimation of the errors in masses, because our MC errors can be too optimistic for masses derived from a limited number of medium resolution spectra with moderate S/N.  
\par
In the future, the red giant will become a low mass white dwarf (WD). Thus TV~Mon will eventually become an EL CVn type eclipsing binary system, like WASP 0346-21, which has similar mass ratio, but shorter period \citep{2024arXiv240717729L}.   

\subsection{Binary evolution simulation}

The low-mass component of TV Mon has a mass of $0.218\pm0.004M_\odot$, which is likely one type of envelope-stripped star, i.e. pre-ELM WD. The formation of (pre-)ELM WD binaries have been well studied in recent years (e.g., \citealt{althaus2013,istrate2016,chenx2017,sunm2018,lizw2019}). Following the work of \citet{lizw2019}, we try to construct the evolutionary history of TV Mon based on the observed parameters. The binary evolution simulations are performed via the detailed stellar evolution code ``Modules for Experiments in Stellar Astrophysics" (MESA, version 12115; \citealt{paxton2011,paxton2013,paxton2015}). We adopt the Ledoux criterion and semi-convection mixing for the convection treatment, where the mixing length parameter is set to be 1.5, and semi-convection is modeled with an efficiency parameter of $\alpha_{\rm sc}=0.01$. We adopted the element abundances of Population I stars, i.e., metallicity $Z=0.02$ and hydrogen mass fraction $X=0.70$. The mass transfer processes are simulated based on the Ritter scheme \citep{ritter1988}. 

We first try to evolve both components simultaneously. Here we assumed that the binary would merge once the accretor fills its Roche lobe. The typical progenitors of pre-ELM WDs are generally in the range of $1.0-2.0M_\odot$ \citep{lizw2019}. According to the total mass of TV Mon, i.e., $\sim 2.3M_\odot$, the progenitor of $M_B$ should be larger than $1.15M_\odot$ (since the initial donor mass is larger than the initial accretor mass). For a given initial donor mass and the chosen accretion efficiency, one can approximately determine the initial accretor mass (initial mass ratio) according to the observational constraints on TV Mon. Then we test a series of binary models with varying initial binary parameters, i.e. donor mass, mass ratio, orbital period, accretion efficiency. Unfortunately, none of these models can reproduce the observation parameters of TV Mon. The binaries would either merge due to the Roche lobe overflow (RLO) from the accretor or produce pre-ELM WDs moremassive than $M_{\rm B}$. Therefore, in the subsequent simulations, we take the accretor as a mass point. 

\begin{figure}
	\includegraphics[width=\columnwidth]{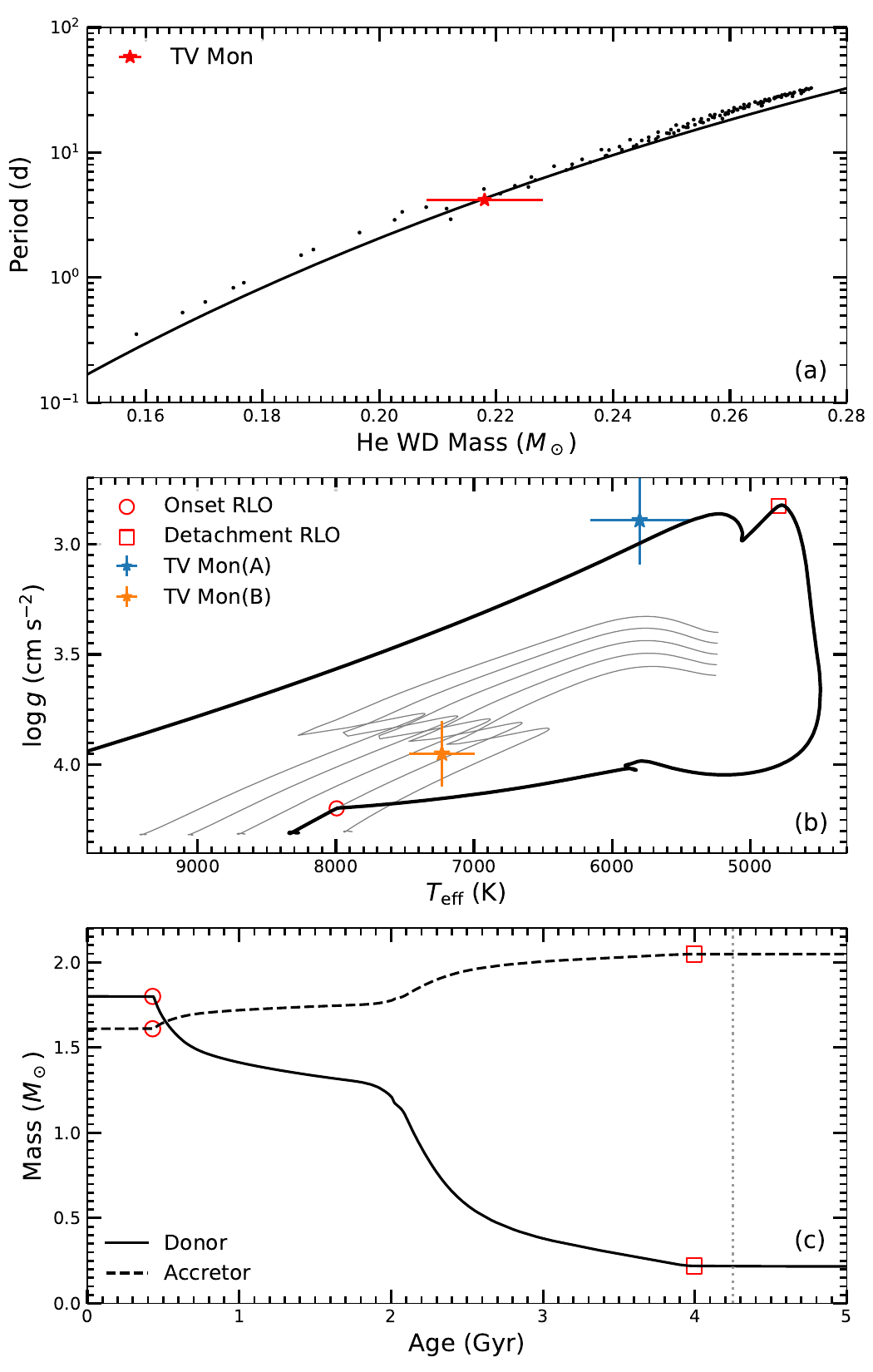}
    \caption{The evolutionary models of TV Mon. Panel (a): The WD mass-orbital period plane in the simulations. The black solid line is for the theoretically fitted curve of \citet{linj2011}. Panel (b): The evolutionary track in $T_{\rm eff}-\log g$ plane for the best model. The black solid line is for the donor in the binary evolution and the grey lines are for the single evolution tracks with masses from $1.7$ to $2.1M_\odot$, from right to left. The observed parameters for the two components of TV Mon are shown in colored stars. The onset and detachment of Roche lobe overflow (RLO) are shown in red open circle and square, respectively. Panel (c): The masses of the two components as a function of the evolutionary age. The grey dotted line is the approximate age in the current state of TV Mon according to the results in panel (a). }
    \label{fig:evo}
\end{figure}

The accretion efficiency is a free parameter and cannot be limited in our simulations. As suggested by our previous work on Algol-type binary \citep{tyc}, the accretion efficiency of $0.3$ can support the observed parameters of TYC 2990-127-1 well. Therefore, our binary simulations adopt a fixed value of $0.3$ for the accretion efficiency. Then the progenitor of $M_{\rm B}$ should be more massive than $1.64M_\odot$. We test four groups of binary models with initial donor masses ranging from $1.7$ to $2.1M_\odot$, the initial accretor masses and orbital separations adjust correspondingly to match the observations. The binary evolution results are shown in Figure~\ref{fig:evo}. In the panel (a), we put the observed and simulated samples in the He WD mass-orbital period plane, where the theoretically fitted curve taken from \citet{linj2011} is also shown for comparison. It is clear the He WD binaries produced from stable RLO processes show a strong correlation between the orbital period and He WD mass. The observational parameters of TV Mon support such a relation pretty well. In panel (b), we present the possible evolutionary history of TV Mon. The thick black line is for the binary evolution model and the thin grey lines are for single evolutionary tracks with masses from $1.7$ to $2.1M_\odot$ (from right to the left), respectively. In our best model, the initial binary contains a $1.8M_\odot$ donor star and a $1.61M_\odot$ accretor, and the initial orbital period is $0.626\;\rm d$. The donor star fills its Roche lobe at the early main sequence stage, as shown in the red open circle. The donor then ascends the red giant branch in the late mass transfer phase. The pre-ELM WD is born after the termination of mass transfer processes, as shown in the red open square. The pre-ELM WD would not enter into the cooling stage immediately because the residual hydrogen layer is still burning, sustaining a relatively high luminosity \citep{istrate2014,chenx2017,lizw2019}. The mass of the pre-ELM WD in panel (b) is $0.218M_\odot$, which coincides with the observations. The accretor mass after the mass transfer is $2.05M_\odot$. The detailed structure of the accretor is not considered in our simulations. According to the single stellar tracks (thin grey lines), we see that the inferred mass of $M_{\rm A}$ is in the range of $\sim 1.7-1.9M_\odot$, which is slighly lower than the observationally derived mass of $M_{\rm A}$ ($2.063\pm 0.033M_\odot$). The discrepancy may originate from the fact that the accretor's rejuvenation process may alter the stellar structure comparing with the same mass single star (e.g. \citealt{zhaoz2024,lau2024}). In panel (c), we present the masses of the two components as a function of evolutionary age. The onset and termination of the mass transfer processes are shown in red open circles and squares, respectively. The age of current state of TV Mon is approximately shown in grey dotted line. We see that the time elapsed since the end of mass transfer is about $250$ Myr. Above all, our simulation could reproduce most of the important observed parameters of TV Mon. Our results suggest that TV Mon contains an envelope-stripped star born via binary interaction. 


\section{Conclusions}
\label{concl}

In this paper we use available observations to characterise dEB system TV~Mon. Here we summarise our results:
\begin{enumerate}
    \item we applied multiple-epochs spectral fitting to LAMOST-MRS data to derive RV and spectral parameters;
    \item we checked the CoRoT LC and found high frequency pulsations of the primary component and signs of long period variability of the secondary component. The analysis of eclipses reveal the relative sizes of the stellar components and edge-on circular orbit;
    \item the combined LC and RV solution provided us estimations for masses and radii of the components: $M_{A,B}=2.063\pm0.033({\rm stat.})\pm0.095({\rm syst.}),~0.218\pm0.004({\rm stat.})\pm0.018({\rm syst.})~M_\odot$, $R_{A,B}=2.394\pm0.014,~2.860\pm0.016~R_\odot$; 
    \item SED analysis and Gaia parallax allowed us to estimate the temperatures ${\teff}_{A,B}=7624^{+194}_{-174},~5184^{+130}_{-123}$ K and distance $d=907\pm11$ pc. 
    \item we measured three $\delta$ Scuti type frequencies in the primary component, while we also suspect TV Mon having a long period variability with period $P_{\rm long}\sim128$ days, although it should be proven by the future long term observations;
    \item this system experienced intensive mass transfer and mass ratio reversal in the past. Now it shows no signs of mass transfer in the spectra. The low mass component will shrink to a He WD, which mass and orbital period are in good agreement with evolutionary models predictions. Eventually it will be EL CVn type eclipsing binary system.
  
\end{enumerate}

TV~Mon is a very interesting laboratory for studying post-mass transfer binary systems. It will benefit from future observations like high-resolution spectroscopy and space-based photometry.  New spectra will be able to confirm our mass estimations with good accuracy. A relatively long (40 min) total eclipse phase allows for a extraction of a detailed spectra of the red giant component, without other component contribution. Future analysis should take into account gravitational red shift and possibly convectional blue shift effects, when new spectra become available. Longer photometrical time series will allow more detailed analysis of $\delta$ Scuti pulsations in the primary and confirmation of possible long period variability of the secondary.


\section*{Acknowledgements}
 We are grateful to the anonymous referee for a constructive report. His suggestions have significantly improved this article. We thank Hans B{\"a}hr (MPIA) for his careful proof-reading of the manuscript.
We thank Hans-Ludwig (LSW) for useful discussions.
MK dedicates this article to Dr. Xiang Liu, no mater if she cares about it or not, work on this paper helped him to pass breakup.

Guoshoujing Telescope (the Large Sky Area Multi-Object Fiber Spectroscopic Telescope LAMOST) is a National Major Scientific Project built by the Chinese Academy of Sciences. Funding for the project has been provided by the National Development and Reform Commission. LAMOST is operated and managed by the National Astronomical Observatories, Chinese Academy of Sciences. 
This work is supported by the National Key R$\&$D Program of China (grant Nos. 2021YFA1600403, 2021YFA1600400), the Natural Science Foundation of China (grant Nos. 12125303, 12288102, 12090040/3, 12473034, 12103086, 12273105, 11703081, 11422324, 12073070), Yunnan Fundamental Research Projects (Nos. 202401BC070007 and 202201BC070003), the Yunnan Revitalization Talent Support Program–Science $\&$ Technology Champion Project (No. 202305AB350003), the Key Research Program of Frontier Sciences of CAS (No. ZDBS-LY-7005), Yunnan Fundamental Research Projects (grant Nos. 202301AT070314, 202101AU070276, 202101AV070001), and the International Centre of Supernovae, Yunnan Key Laboratory (No. 202302AN360001). We also acknowledge the science research grant from the China Manned Space Project with Nos. CMS-CSST-2021-A10 and CMS-CSST-2021-A08. The authors gratefully acknowledge the “PHOENIX Supercomputing Platform” jointly operated by the Binary Population Synthesis Group and the Stellar Astrophysics Group at Yunnan Observatories, Chinese Academy of Sciences. 
This research has made use of NASA’s Astrophysics Data System, the SIMBAD data base, and the VizieR catalogue access tool, operated at CDS, Strasbourg, France. It also made use of TOPCAT, an interactive graphical viewer and editor for tabular data \citep[][]{topcat}. 
{\it Gaia} (\url{https://www.cosmos.esa.int/gaia}), processed by the {\it Gaia} Data Processing and Analysis Consortium (DPAC, \url{https://www.cosmos.esa.int/web/gaia/dpac/consortium}). Funding for the DPAC has been provided by national institutions, in particular the institutions participating in the {\it Gaia} Multilateral Agreement.
This research has made use of the NASA/IPAC Infrared Science Archive, which is funded by the National Aeronautics and Space Administration and operated by the California Institute of Technology.

\section*{Data Availability}
The data underlying this article will be shared on reasonable request to the corresponding author.




\bibliographystyle{mnras}



\appendix

\section{Spectral models}
\label{sec:payne}
The synthetic spectra are generated using NLTE~MPIA online-interface \url{https://nlte.mpia.de} \citep[see Chapter~4 in][]{disser} on wavelength intervals 4870:5430 \AA~for the blue arm and 6200:6900 \AA ~for the red arm with spectral resolution $R=7500$. We use NLTE (non-local thermodynamic equilibrium) spectral synthesis for H, Mg~I, Si~I, Ca~I, Ti~I, Fe~I and Fe~II lines \citep[see Chapter~4 in][ for references]{disser}.  
\par
The grid of models (6200 in total) is computed for points randomly selected in a range of $\teff$ between 4600 and 8800 K, $\logg$ between 1.0 and 4.8  (cgs units), $\vsini$ from 1 to 300 $\kms$ and [Fe/H]\footnote{We used $\feh$ as a proxy of overall metallicity, abundances for all elements are scaled with Fe.} between $-$0.9 and $+$0.9 dex. The model is computed only if linear interpolation of the MAFAGS-OS\citep[][]{Grupp2004a,Grupp2004b} stellar atmosphere is possible for a given point in parameter space.  Microturbulence is fixed to $\Vmic=2~\kms$ for all models. 

\section{WISE light curves}
\label{sec:wise}
\begin{figure}
	\includegraphics[width=\columnwidth]{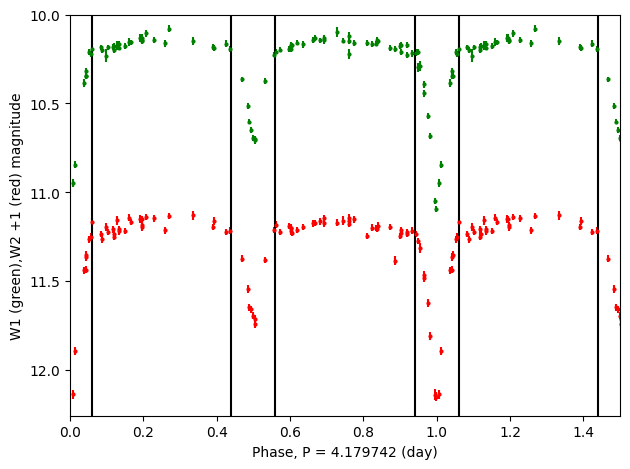}
	\caption{{\sc WISE} light curves.}
    \label{fig:wise}
\end{figure}

We show {\sc WISE} light curves in Figure~\ref{fig:wise}. Eclipses (parts between black vertical lines) were excluded, when we recompute photometry for SED fitting.

\section{RV measurements}
 We provide RV measurements from the multiple-epoch fit in Table~\ref{tab:rvs}.
\begin{table}
    \centering
    \caption{\label{tab:rvs} Radial velocity measurements. Errors were adjusted by {\sc JKTEBOP} to reach unity in reduced $\chi^2$. We subtract 2400000.5 days from time values.}
    \begin{tabular}{lccc}
\hline
BJD & RV$_{A}$ & RV$_{B}$& phase \\
 d  & $\kms$ & $\kms$ &  \\
 \hline
  58530.5432&    40.93$\pm$ 0.16& -55.73$\pm$0.90 &   0.593\\
  58530.5599&    41.13$\pm$ 0.17& -57.64$\pm$1.05 &   0.597\\
  58530.5828&    41.70$\pm$ 0.17& -63.03$\pm$1.10 &   0.606\\
  58850.6423&    16.83$\pm$ 0.10& 172.54$\pm$0.79 &   0.176\\
  58850.6562&    16.82$\pm$ 0.10& 172.60$\pm$0.82 &   0.180\\
  58850.6687&    16.55$\pm$ 0.10& 175.15$\pm$0.75 &   0.183\\
  58850.6882&    16.38$\pm$ 0.11& 176.79$\pm$0.77 &   0.187\\
  58850.7014&    16.22$\pm$ 0.10& 178.30$\pm$0.75 &   0.191\\
  58850.7139&    16.06$\pm$ 0.11& 179.75$\pm$0.88 &   0.194\\
  58850.7319&    15.89$\pm$ 0.11& 181.41$\pm$1.01 &   0.198\\
  58850.7444&    15.63$\pm$ 0.11& 183.82$\pm$1.01 &   0.201\\
  58909.5526&    15.40$\pm$ 0.16& 186.05$\pm$1.56 &   0.271\\
  58909.5658&    15.44$\pm$ 0.16& 185.62$\pm$1.65 &   0.274\\
  58909.5789&    15.49$\pm$ 0.17& 185.20$\pm$1.78 &   0.277\\
  58911.5517&    48.45$\pm$ 0.24& -126.91$\pm$1.78 &   0.749\\
  58911.5649&    48.45$\pm$ 0.24& -126.92$\pm$1.82 &   0.752\\
  58911.5774&    48.26$\pm$ 0.26& -125.12$\pm$1.91 &   0.755\\
  58913.4793&    15.24$\pm$ 0.44& 187.62$\pm$5.32 &   0.210\\
  58913.4918&    15.52$\pm$ 0.40& 184.90$\pm$4.73 &   0.213\\
  58913.5147&    15.71$\pm$ 0.33& 183.13$\pm$3.91 &   0.219\\
  58913.5328&    15.63$\pm$ 0.30& 183.83$\pm$3.41 &   0.223\\
  58913.5453&    15.00$\pm$ 0.27& 183.83$\pm$3.19 &   0.226\\
  58913.5585&    14.97$\pm$ 0.31& 190.22$\pm$3.69 &   0.229\\
  \hline
 \end{tabular}
 \end{table}

\section{TLCM modelling}
\label{sec:tlcm}
The presence of a total eclipse and transit in the CoRoT LC allows us to use {\sc TLCM}, which was developed and actively used to fit exoplanet transits in CoRoT data \citep{tlcm}. It allows for the use of RV data from two components in ``SB2" mode and estimate parameters uncertainties using Monte Carlo simulations. Reflection, ellipsoidal and beaming effects are included in modelling for both components. We fix period $P=4.179741$ day and eccentricity $e=0$, while fitting for all other parameters, including limb darkening coefficients for both stars $u_{+,-}$ and parabolic trend with time. The final solution is determined as a median of 25000 MC simulations with 10 walkers. 
\par
\begin{figure}
	\includegraphics[width=\columnwidth]{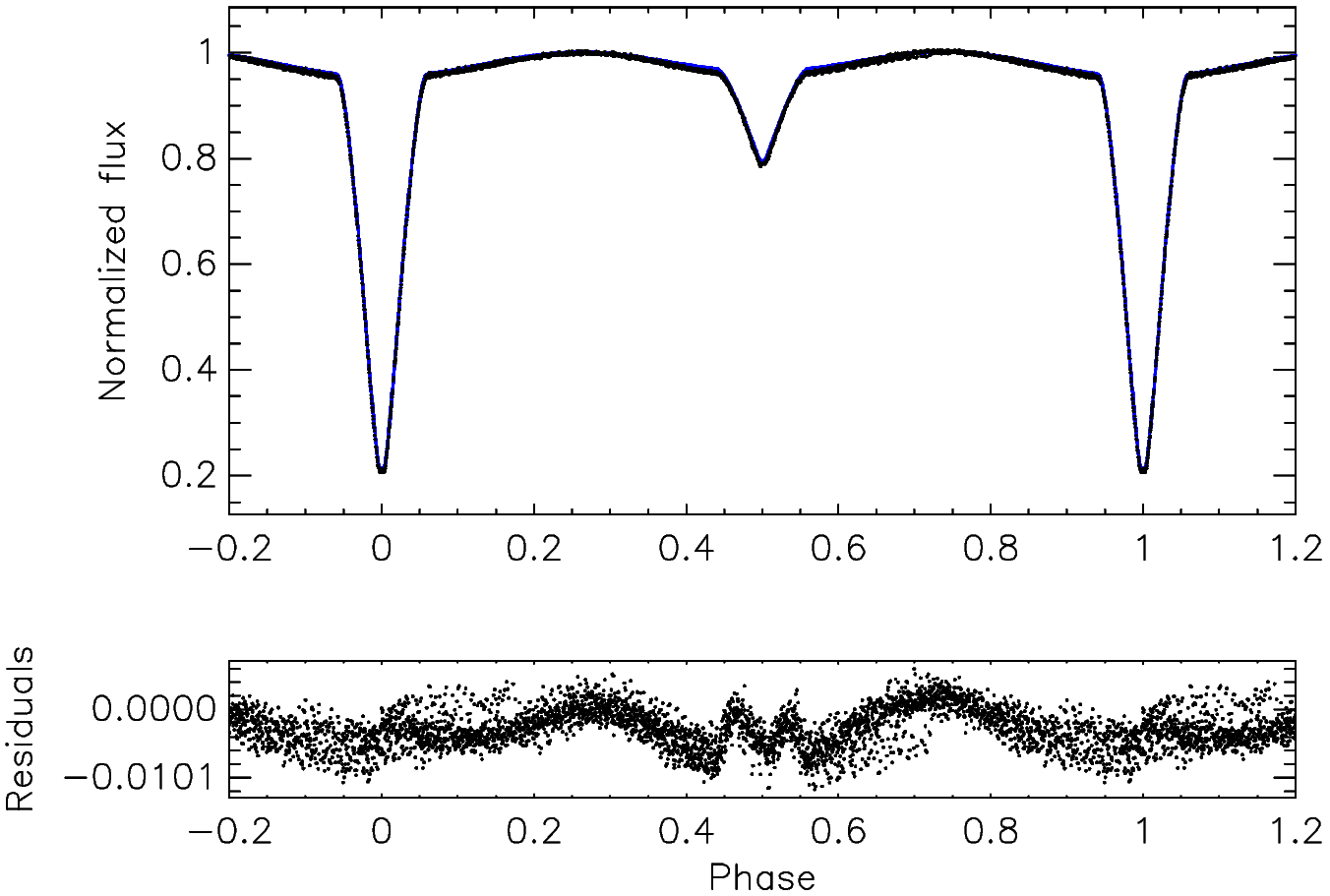}
    \includegraphics[width=\columnwidth]{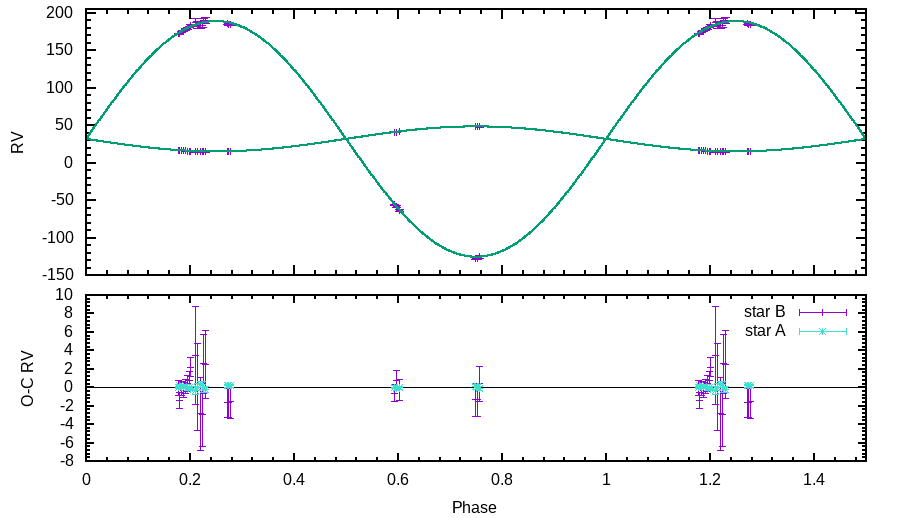}
    \caption{Phase-folded LC (top) and RV (bottom) fits with {\sc TLCM}. The periodic structure is obvious in the LC residuals around phase=0.5.}
    \label{fig:tlcm}
\end{figure}

Results derived by {\sc TLCM} are collected in Table~\ref{tab:tlcm} and shown in Figure~\ref{fig:tlcm}. They are consistent with the {\sc JKTEBOP} solution, although slightly different. The total eclipse is well fitted, while for the secondary the minimum residuals are symmetric relative to phase 0.5. This code was designed to fit exoplanet transits, so it can be expected to perform worse for stars. Anyway residuals are comparable to the amplitude of high-frequency pulsations. The out-of-eclipse part of the LC is dominated by ellipsoidal variability (Ell.$_{A,B}^{\rm max}\sim 0.25,~6.0$ per cent of total light) and reflection (Refl.$_{A,B}^{\rm max}\sim0.3,~1.7$ per cent of total light), while the light beaming effect is negligible (Beam.$_{A,B}^{\rm max}\sim 0.016,~0.06$ per cent of total light).   

\begin{table}
    \centering
        \caption{{\sc TLCM} solution. Error estimates are from 25000 MC simulations.}
    \begin{tabular}{l|c}
\hline
Parameter & value\\
\hline
fixed:\\
${\teff}_A$, K& {7232}\\
$\sqrt{e}\cos{\omega}$  &  {0.0}\\
$\sqrt{e}\sin{\omega}$  &  {0.0}\\
$P$, d  & 4.179741\\
\hline
fitted:\\
$J$   & 0.2281$\pm$0.0003\\
$a/R_A$ & 5.776$\pm$0.012\\
$R_B/R_A$  & 1.1027$\pm$0.0024\\
Conjunction parameter $b$& $0.000072\pm0.014472$\\
$u_{+,-}$ A & 0.86$\pm$0.03,~$-0.59\pm0.12$\\
$u_{+,-}$ B & 0.05$\pm$0.02,~$-1.01\pm0.08$\\
Albedo A & 0.63$\pm$0.14 \\
Albedo B & 0.66$\pm$0.03 \\
$t_0$, BJD d& 2454457.49559$\pm$0.00006  \\
$K_A,\,\kms$  & 16.634$\pm$0.065 \\
$q$ & 0.1057$\pm$0.0004\\
$\gamma,\,\kms$ & 31.68$\pm$0.06 \\

\hline
derived\\%
$R_A/a$  &0.1731$\pm$0.0004 \\
${\teff}_B$, K& 5177$\pm$2\\
$i^{\circ}$  & 89.999$^{+0.007}_{-0.141}$\\
$a,\,R_\odot$      & 14.375$\pm$0.076 \\
$M_A,\,M_\odot$    & 2.063$\pm$0.033 \\
$M_B,\,M_\odot$    & 0.218$\pm$0.003  \\
$R_A,\,R_\odot$    & 2.486$\pm$0.014 \\
$R_B,\,R_\odot$    & 2.742$\pm$0.017 \\
$\logg_A$, cgs     & 3.953 \\%
$\logg_B$, cgs     & 2.893 \\
\hline
    \end{tabular}
    \label{tab:tlcm}
\end{table}

\section{{\sc emcee} sample of {\sc ellc} solution}

\begin{figure*}
    \includegraphics[width=\textwidth]{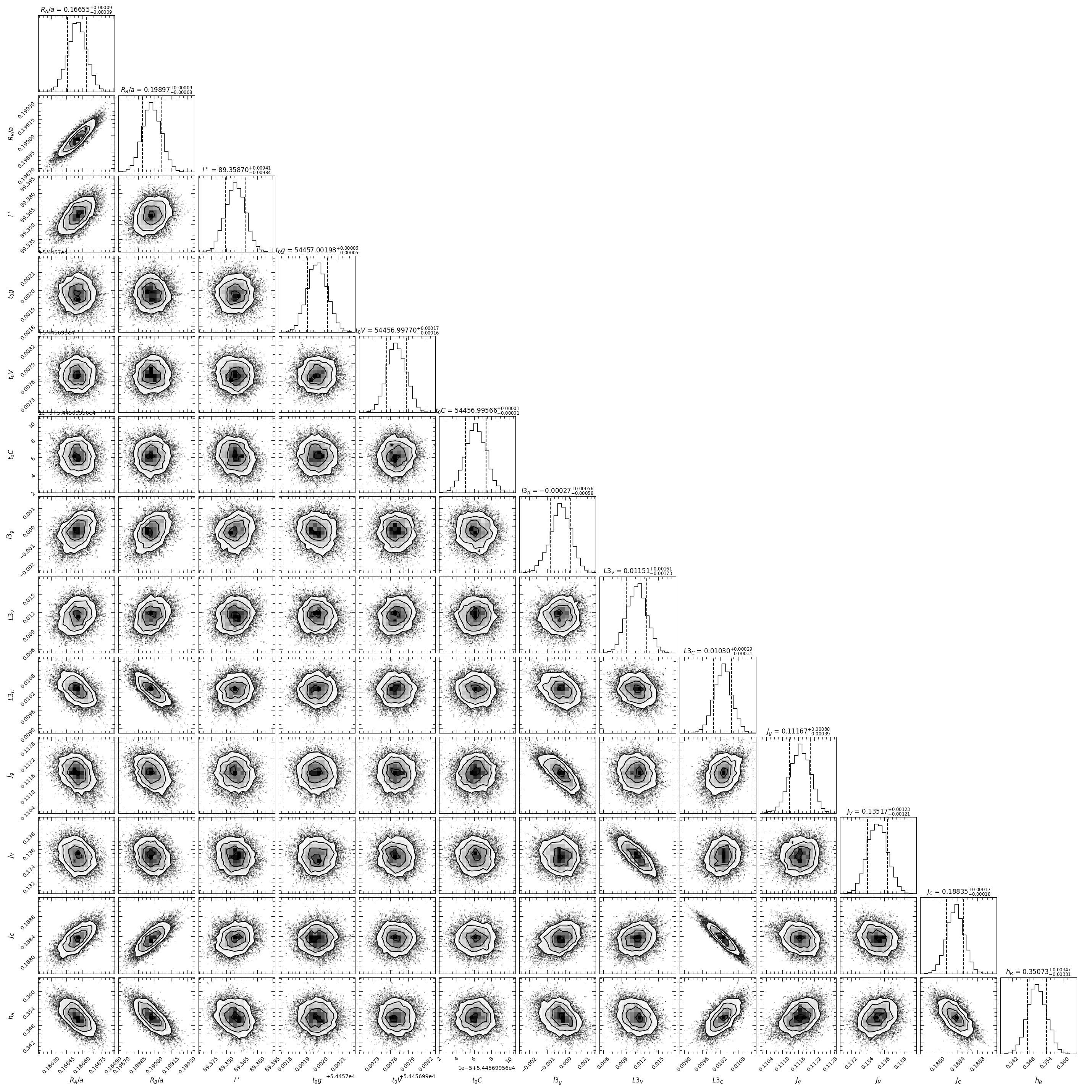}
    \caption{Corner plot for the sampling of the {\sc ellc} solution. Titles show 16, 50 and 84 percentiles.}
    \label{fig:corner}
\end{figure*}

In Figure~\ref{fig:corner} we show corner plot \citep{corner} with {\sc emcee} sampling results for {\sc ellc} solution.

\section{Full datasets with {\sc W-D}}
\label{sec:fullwd}
\begin{figure}
    \centering
    \includegraphics[width=0.95\columnwidth]{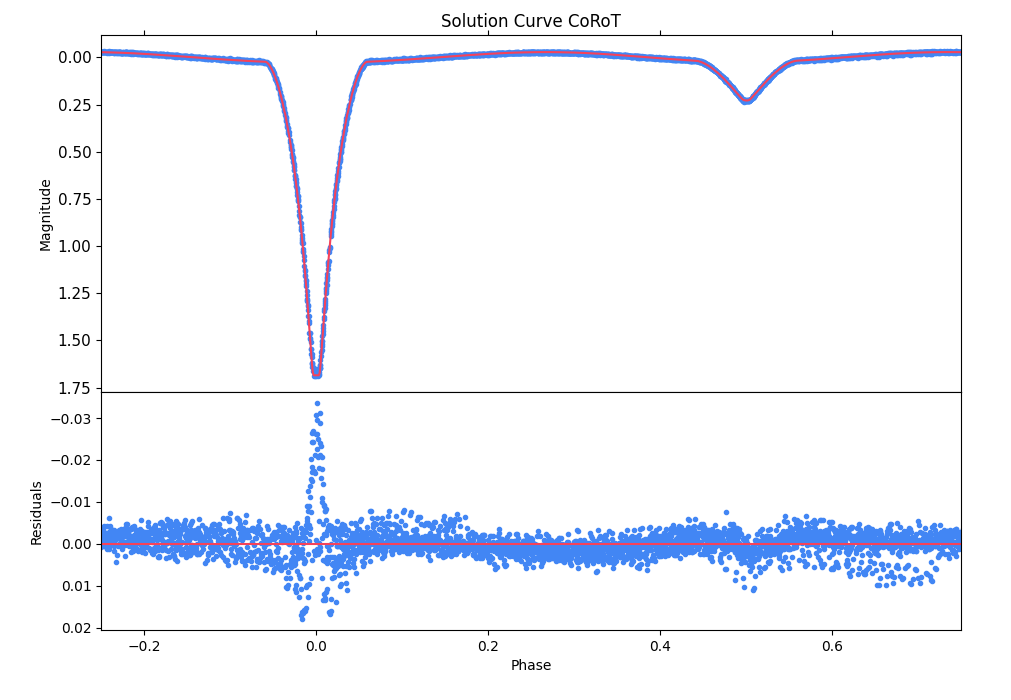}
    \includegraphics[width=0.95\columnwidth]{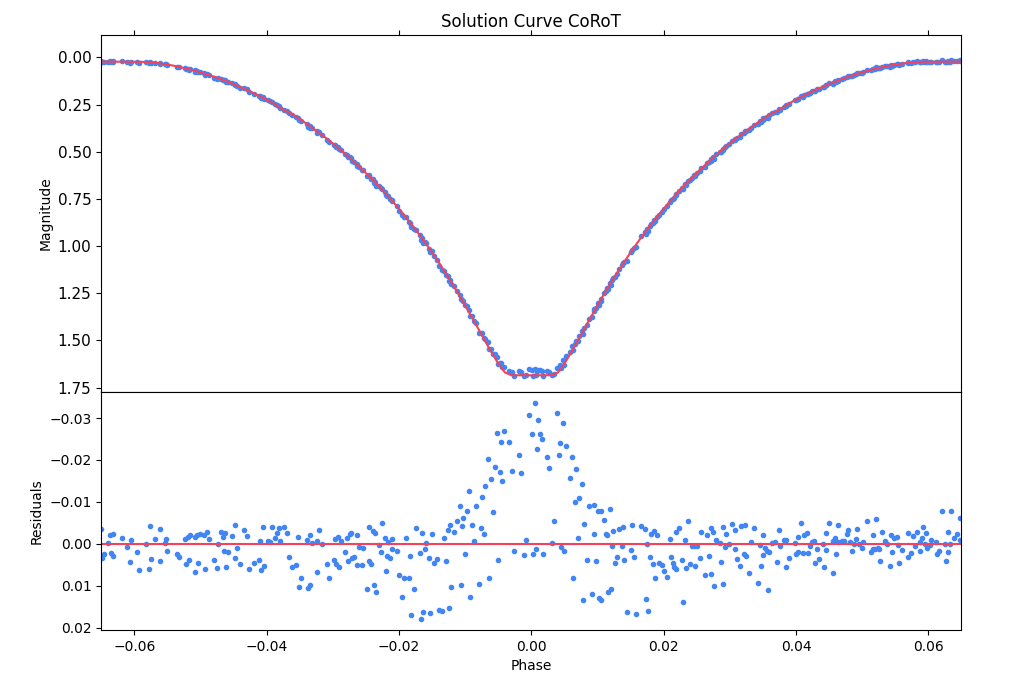}
    \includegraphics[width=0.95\columnwidth]{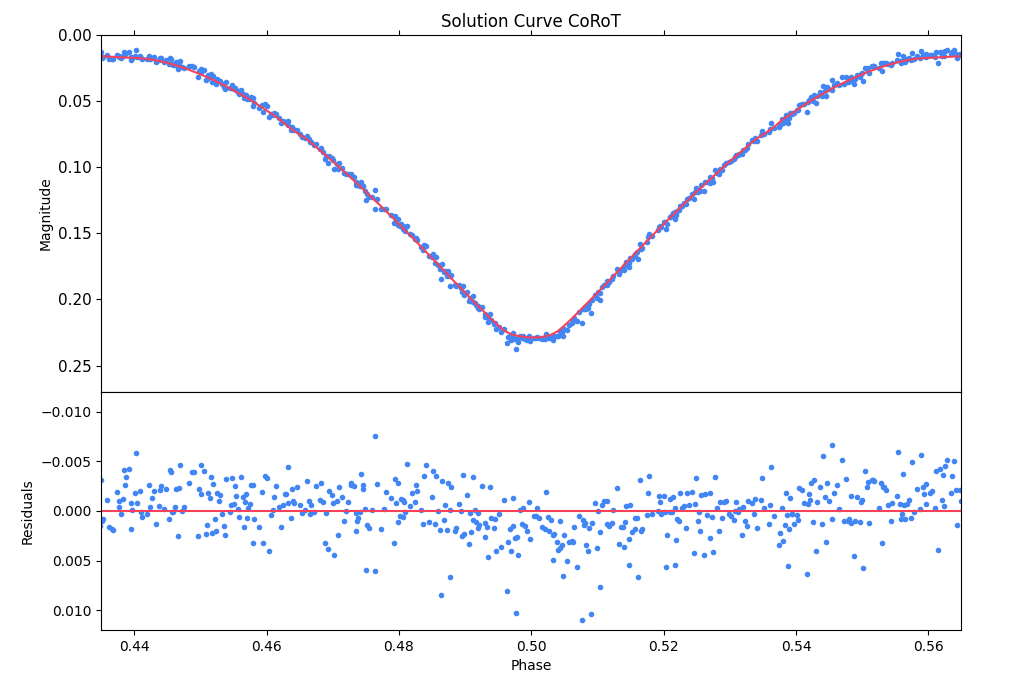}
    \caption{The LC solution by {\sc W-D} using full datasets (CoRoT). Top panels show fit of the data, bottom panels show fit residuals. }
    \label{fig:pywd0}
\end{figure}

\begin{figure}
    \centering
    \includegraphics[width=0.95\columnwidth]{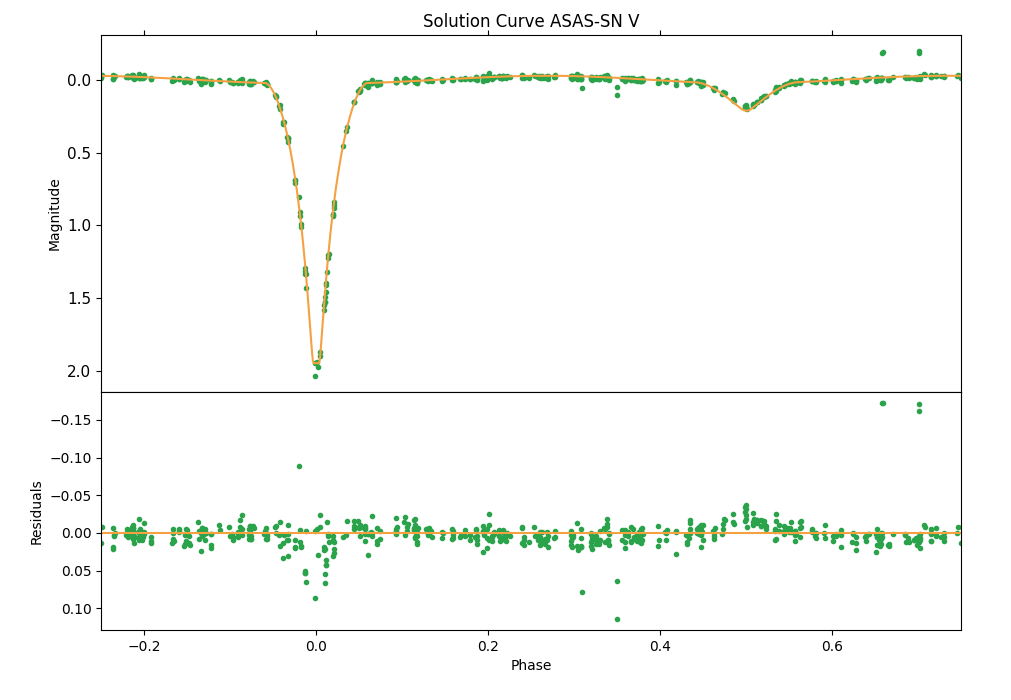}
    \includegraphics[width=0.95\columnwidth]{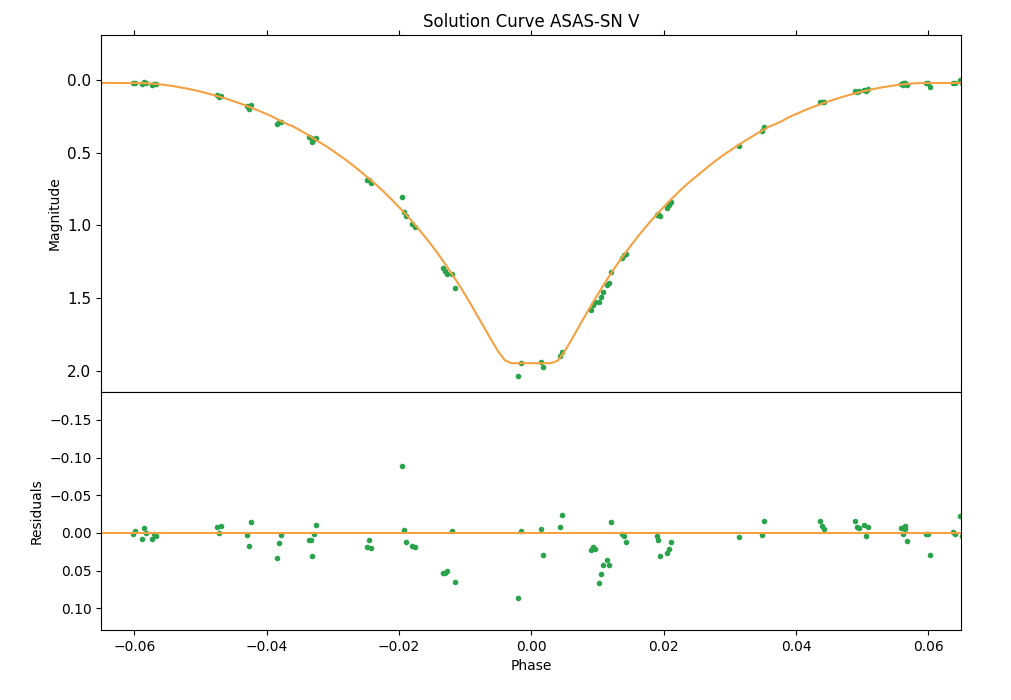}
    \includegraphics[width=0.95\columnwidth]{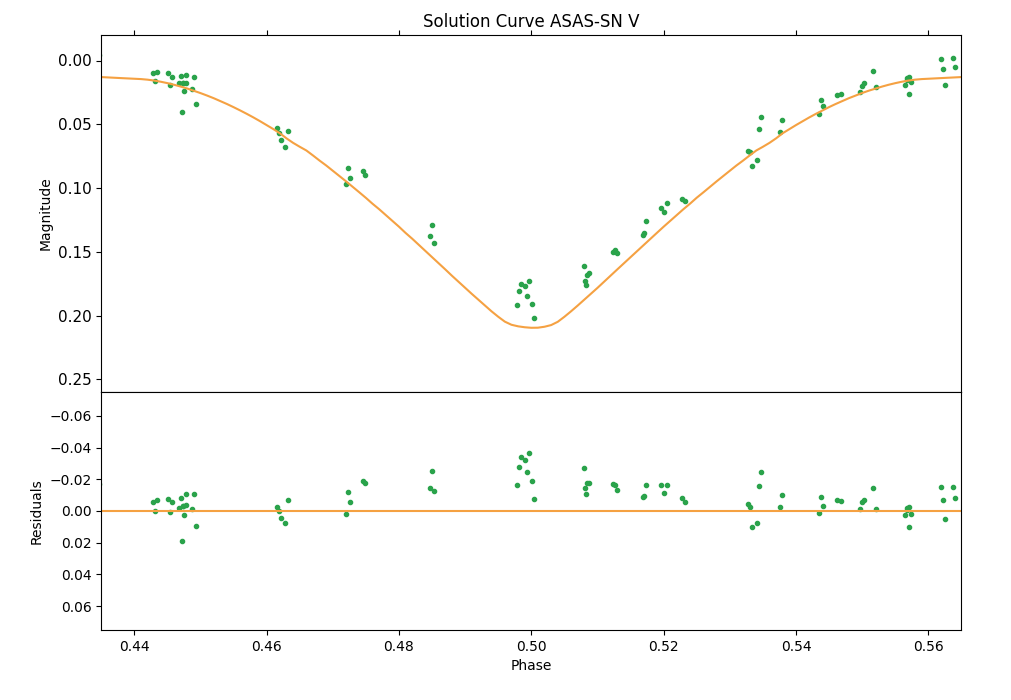}
    \caption{The LC solution by {\sc W-D} using full datasets (ASAS-SN $V$). Top panels show fit of the data, bottom panels show fit residuals. }
    \label{fig:pywd1}
\end{figure}

\begin{figure}
    \centering
    \includegraphics[width=0.95\columnwidth]{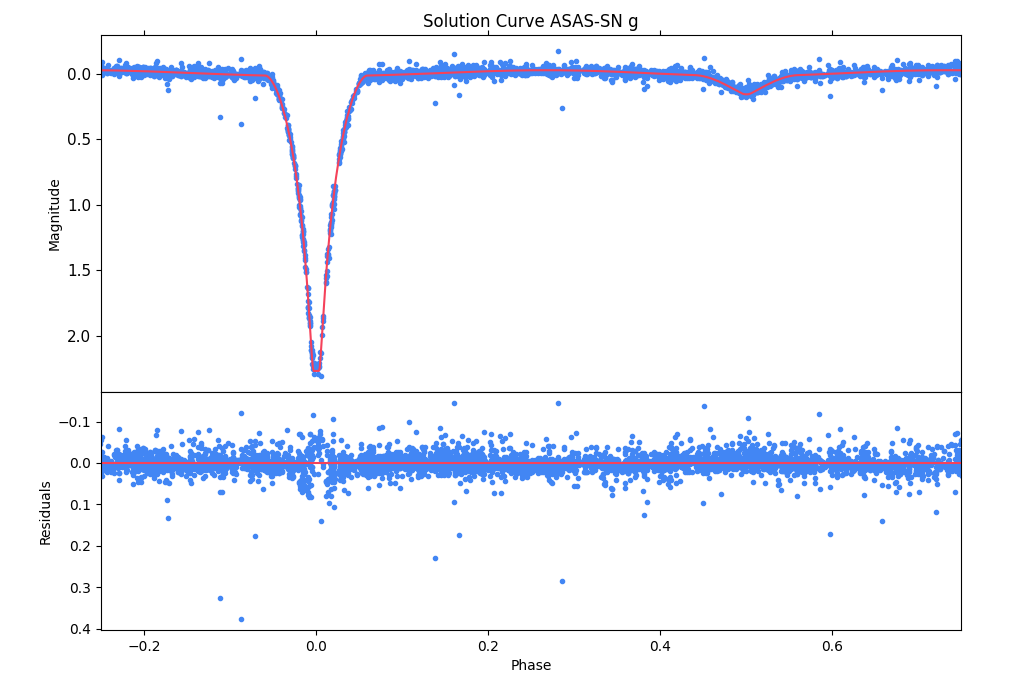}
    \includegraphics[width=0.95\columnwidth]{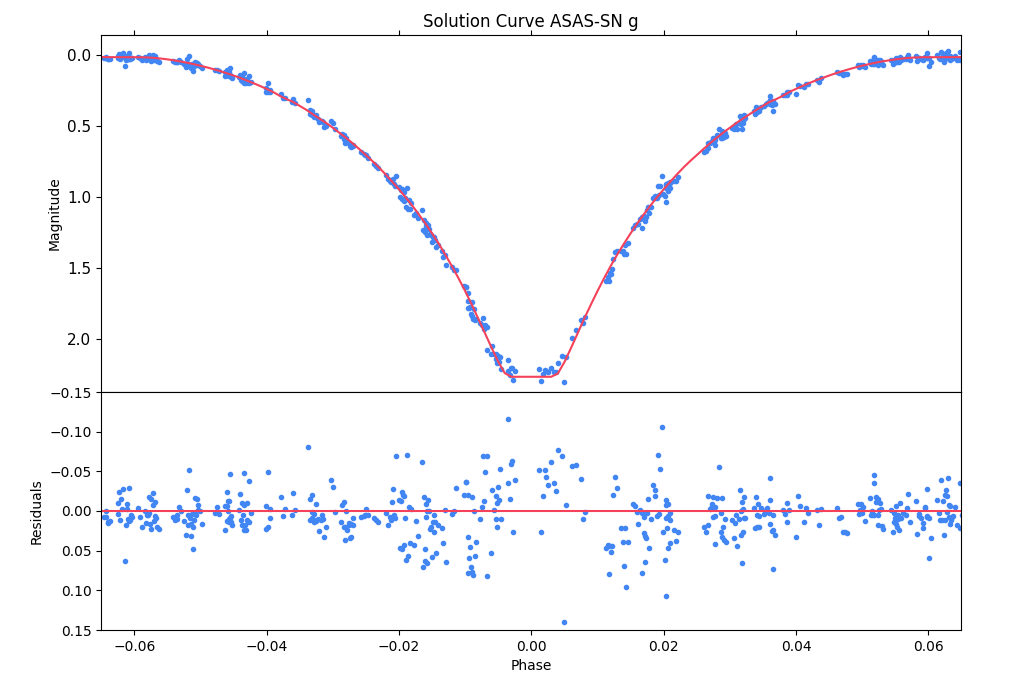}
    \includegraphics[width=0.95\columnwidth]{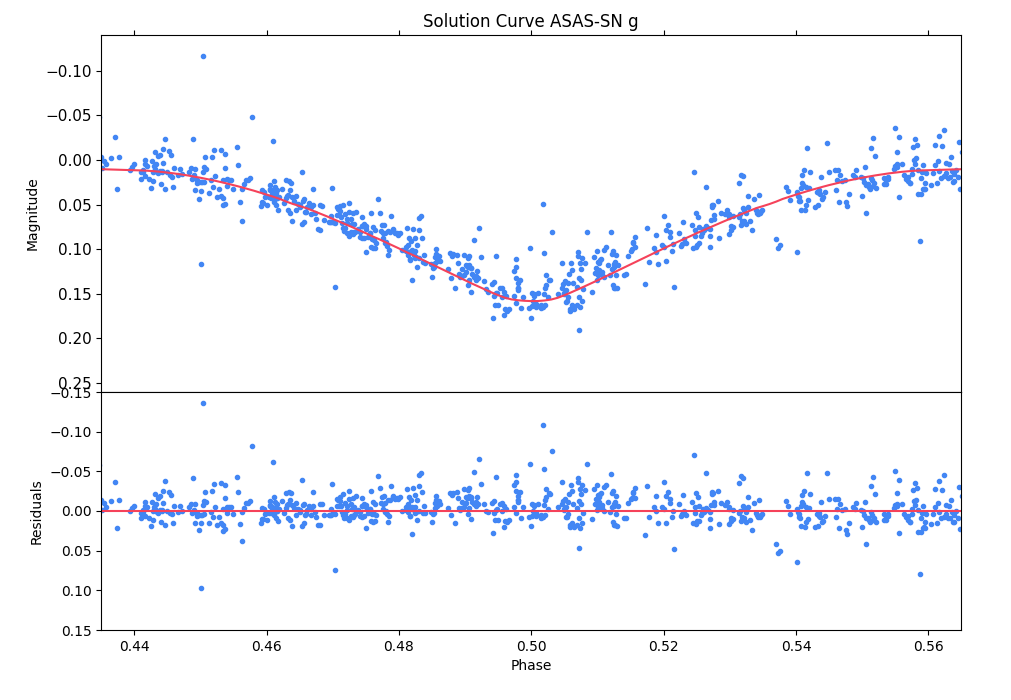}
    \caption{The LC solution by {\sc W-D} using full datasets (ASAS-SN $g$). Top panels show fit of the data, bottom panels show fit residuals. }
    \label{fig:pywd2}
\end{figure}

 We show the solution derived after 21 iterations using three full datasets in Figures~\ref{fig:pywd0},~\ref{fig:pywd1},~\ref{fig:pywd2} and Table~\ref{tab:pywd}. Big residuals in the total eclipse are from the part of the CoRoT LC, which was not used during previous analysis by the {\sc ellc} and {\sc W-D} codes. In comparison with Figure~\ref{fig:ecl_ellc} the out-eclipse part is fitted well by {\sc W-D}. Both ASAS-SN LCs are also fitted well, although for $V$ band secondary eclipse is slightly shallower in comparison with the best fit model.

\begin{table}
    \centering
        \caption{{\sc W-D} solution for full three datasets.}
    \begin{tabular}{l|c}
\hline
Parameter & value\\
\hline
fixed:\\
$P$, d  & 4.179741\\
$a,\,R_\odot$ & 14.375 \\
$q$ & 0.1056\\
$e$ & 0.0\\
$A_A$ &1.0\\
${\teff}_A$, K & $7232$\\
\hline
fitted:\\
${\teff}_B$, K & $5067\pm1$\\
$i^{\circ}$  & 89.87$\pm$0.08\\
$L1_g$   & 10.732$\pm$0.025\\
$L1_V$   & 10.139$\pm$0.056\\
$L1_C$   & 9.528$\pm$0.006\\
$L3_g$   & -0.005$\pm$0.002\\
$L3_V$   & -0.005$\pm$0.005\\
$L3_C$   & 0.026$\pm$0.001\\
$A_B$ & 0.84$\pm$0.01 \\
$\Omega_A$ & $6.027\pm0.005$\\
$\Omega_B$ & $2.001\pm0.001$\\
\hline
derived\\%
$r_1$ (pole) & 0.16883$\pm$0.00014 \\
$r_1$ (point) & 0.16943$\pm$0.00014 \\
$r_1$ (side) & 0.16928$\pm$0.00014 \\
$r_1$ (back) & 0.16940$\pm$0.00014 \\

$r_2$ (pole) & 0.185075$\pm$0.000005 \\
$r_2$ (point) & 0.233134$\pm$0.00016 \\
$r_2$ (side) & 0.191494$\pm$0.000006 \\
$r_2$ (back) & 0.215513$\pm$0.00001 \\

$M_A,\,M_\odot$ & 2.063 \\
$M_B,\,M_\odot$ & 0.218 \\
$K_A,\,\kms$ & 16.619 \\
$K_B,\,\kms$ & 157.379 \\
$R_A,\,R_\odot$ & 2.431 \\
$R_B,\,R_\odot$ & 2.838 \\
$\log{L_A},\,L_\odot$ & 1.161 \\
$\log{L_B},\,L_\odot$ & 0.678 \\
$L_B/L_A$ & 0.293\\
$\logg_A$, cgs  & 3.98 \\%
$\logg_B$, cgs  & 2.87 \\
\hline
    \end{tabular}
    \label{tab:pywd}
\end{table}

\bsp	
\label{lastpage}
\end{document}